\documentclass[nopacs,twocolumn,epjc3]{svjour3}
\pdfoutput=1
\usepackage[preprint]{atlascover}
\usepackage{graphicx}
\usepackage{overpic}
\usepackage{array}
\usepackage{subfigure}

\usepackage{placeins}
\usepackage{booktabs}
\usepackage{xspace}
\usepackage[switch]{lineno}
\usepackage{cite}
\usepackage{hyphenat}
\usepackage{amsmath}
\usepackage{hyperref}
\hypersetup{colorlinks,citecolor=blue,filecolor=blue,linkcolor=blue,urlcolor=blue}

\newcommand{\bbbar}{\ensuremath{b\bar{b}}}
\newcommand{\ttbar}{\ensuremath{t\bar{t}}}
\newcommand{\ifb}{\ensuremath{\mathrm{fb}^{-1}}}
\newcommand{\chinopm}{\ensuremath{\tilde{\chi}^\pm}}

\newcommand{\chinoonepm}{\ensuremath{\tilde{\chi}_1^\pm}}
\newcommand{\chinopmonetwo}{\ensuremath{\tilde{\chi}_{1,2}^\pm}}

\newcommand{\ninoone}{\ensuremath{\tilde{\chi}_1^0}}
\newcommand{\ninotwo}{\ensuremath{\tilde{\chi}_2^0}}

\newcommand{\ninoonetwothreefour}{\ensuremath{\tilde{\chi}_{1,2,3,4}^0}}

\newcommand{\pt}{\ensuremath{p_\mathrm{T}}}
\newcommand{\met}{\ensuremath{E_{\mathrm{T}}^{\mathrm{miss}}}}
\newcommand{\metrel}{\ensuremath{E_{\mathrm{T}}^{\mathrm{miss,rel}}}}

\newcommand{\meff}{\ensuremath{m_\mathrm{eff}}}
\newcommand{\mtmax}{\ensuremath{m_\mathrm{T}^\mathrm{max}}}
\newcommand{\mlj}{\ensuremath{m_{\ell j}}}
\newcommand{\mljj}{\ensuremath{m_{\ell jj}}}

\newcommand{\lumi}{20.3\,\ifb}
\newcommand{\lumierror}{2.8\%}

\newcommand{\comblimit}{250\,GeV}

\def\TeV{\ifmmode {\mathrm{\ Te\kern -0.1em V}}\else \textrm{Te\kern -0.1em V}\fi}
\def\GeV{\ifmmode {\mathrm{\ Ge\kern -0.1em V}}\else \textrm{Ge\kern -0.1em V}\fi}
\def\MeV{\ifmmode {\mathrm{\ Me\kern -0.1em V}}\else \textrm{Me\kern -0.1em V}\fi}
\def\keV{\ifmmode {\mathrm{\ ke\kern -0.1em V}}\else \textrm{ke\kern -0.1em V}\fi}
\def\eV{\ifmmode  {\mathrm{\ e\kern -0.1em V}}\else \textrm{e\kern -0.1em V}\fi}

\newcommand{\mNone}{\ensuremath{m_{\smash{\ninoone}}}}

\newcommand{\alpgen}{{\textsc{Alpgen}}\xspace}

\newcommand{\mcatnlo}{{\textsc{MC@NLO}}\xspace}

\newcommand{\acermc}{{\textsc{AcerMC}}\xspace}
\newcommand{\sherpa}{{\textsc{Sherpa}}\xspace}
\newcommand{\herwig}{{\textsc{Herwig}}\xspace}
\newcommand{\herwigpp}{{\textsc{Herwig{\tt++}}}\xspace}

\newcommand{\prospino}{{\textsc{Prospino2}}\xspace}
\newcommand{\madgraph}{{\textsc{MadGraph}}\xspace}

\newcommand{\cteqsixl}{{\textsc{CTEQ6L1}}\xspace}

\newcommand{\ctten}{{\textsc{CT10}}\xspace}
\newcommand{\pythia}{{\textsc{Pythia6}}\xspace}
\newcommand{\pythiaeight}{{\textsc{Pythia8}}\xspace}

\newcommand{\geant}{{\textsc{GEANT4}}\xspace}

\newcommand{\powheg}{{\textsc{Powheg}}\xspace}
\newcommand{\powhegpythia}{\textsc{Powheg}+\textsc{Pythia6}}
\newcommand{\perugia}{{\textsc{Perugia2011C}}\xspace}
\newcommand{\auet}{{\textsc{AUET2B}}\xspace}

\newcommand{\au}{{\textsc{AU2}}\xspace}

\newcommand{\histfitter}{{\textsc{HistFitter}}}

\newcommand{\tth}{\ensuremath{\ttbar h}}
\newcommand{\wjets}{\ensuremath{W+\mathrm{jets}}}
\newcommand{\zjets}{\ensuremath{Z+\mathrm{jets}}}
\newcommand{\mtW}{\ensuremath{m_\mathrm{T}^W}}

\newcommand{\pTvec}{\vec{p}_\mathrm{T}}

\newcommand{\pTmiss}{\vec{p}_\mathrm{T}^\mathrm{\,miss}}
\newcommand{\mct}{\ensuremath{m_\mathrm{CT}}\xspace}

\newcommand{\mbj}{\ensuremath{m_{bj}}\xspace}
\newcommand{\mbb}{\ensuremath{m_{bb}}\xspace}

\newcommand{\nlep}{\ensuremath{n_\mathrm{lepton}}}
\newcommand{\ngamma}{\ensuremath{n_{\gamma}}}
\newcommand{\njet}{\ensuremath{n_\mathrm{jet}}}
\mathchardef\mhyphen="2D
\newcommand{\nbjet}{\ensuremath{n_{b\mhyphen\mathrm{jet}}}}
\newcommand{\mgg}{\ensuremath{m_{\gamma\gamma}}}
\newcommand{\mtgone}{\ensuremath{m_{\mathrm{T}}^{\smash{W\!\gamma_1}}}}
\newcommand{\mtgtwo}{\ensuremath{m_{\mathrm{T}}^{\smash{W\!\gamma_2}}}}
\newcommand{\DphiWh}{\ensuremath{\Delta\phi(W,h)}}
\newcommand{\etcone}[1]{\ensuremath{E_\mathrm{T}^{\mathrm{cone}#1}}}
\newcommand{\ptcone}[1]{\ensuremath{p_\mathrm{T}^{\mathrm{cone}#1}}}
\newcommand{\et}{\ensuremath{E_\mathrm{T}}}
\newcommand{\jvf}{\ensuremath\mathrm{JVF}}
\newcommand{\mll}{\ensuremath{m_{\ell\ell}}}
\newcommand{\detall}{\ensuremath{\Delta\eta_{\ell\ell}}}

\newcommand{\limvisobs}{\ensuremath{\langle\sigma_{\rm vis}\rangle_{\mathrm{obs}}^{95}}}
\newcommand{\limsobs}{\ensuremath{S_\mathrm{obs}^{95}}}
\newcommand{\limsexp}{\ensuremath{S_\mathrm{exp}^{95}}}
\newcommand{\clb}{\ensuremath{CL_B}}

\newcommand{\CLs}{CL$_\mathrm{s}$}
\newcommand{\pzero}{\ensuremath{p_0}}

\newcommand{\lbb}{\ensuremath{\ell{}bb}}
\newcommand{\lgg}{\ensuremath{\ell\gamma\gamma}}
\newcommand{\llss}{\ensuremath{\ell^{\pm}\ell^{\pm}}}
\newcommand{\SRlbbone}{SR$\ell bb$\nobreakdash-\hspace{0pt}1}
\newcommand{\SRlbbtwo}{SR$\ell bb$\nobreakdash-\hspace{0pt}2}
\newcommand{\CRlbbT}{CR$\ell bb$\nobreakdash-\hspace{0pt}T}
\newcommand{\CRlbbW}{CR$\ell bb$\nobreakdash-\hspace{0pt}W}
\newcommand{\VRlbbone}{VR$\ell bb$\nobreakdash-\hspace{0pt}1}
\newcommand{\VRlbbtwo}{VR$\ell bb$\nobreakdash-\hspace{0pt}2}
\newcommand{\SRlggone}{SR$\ell\gamma\gamma$\nobreakdash-\hspace{0pt}1}
\newcommand{\SRlggtwo}{SR$\ell\gamma\gamma$\nobreakdash-\hspace{0pt}2}
\newcommand{\VRlggone}{VR$\ell\gamma\gamma$\nobreakdash-\hspace{0pt}1}
\newcommand{\VRlggtwo}{VR$\ell\gamma\gamma$\nobreakdash-\hspace{0pt}2}
\newcommand{\SReeone}{SR$ee$\nobreakdash-\hspace{0pt}1}
\newcommand{\SReetwo}{SR$ee$\nobreakdash-\hspace{0pt}2}
\newcommand{\SRmmone}{SR$\mu\mu$\nobreakdash-\hspace{0pt}1}
\newcommand{\SRmmtwo}{SR$\mu\mu$\nobreakdash-\hspace{0pt}2}
\newcommand{\SRemone}{SR$e\mu$\nobreakdash-\hspace{0pt}1}
\newcommand{\SRemtwo}{SR$e\mu$\nobreakdash-\hspace{0pt}2}
\newcommand{\SRllone}{SR$\ell\ell$\nobreakdash-\hspace{0pt}1}
\newcommand{\SRlltwo}{SR$\ell\ell$\nobreakdash-\hspace{0pt}2}

\def\thetitle{Search for direct pair production of a chargino and a neutralino decaying to the 125~GeV Higgs boson in \boldmath$\sqrt{s}$~=~8~TeV \boldmath$pp$ collisions with the ATLAS detector}

\def\theabstract{
\begin{sloppypar}
A search is presented for the direct pair production of a chargino and a neutralino $pp\to\chinoonepm\ninotwo$, where the chargino decays to the lightest neutralino and the $W$ boson, $\chinoonepm\to \ninoone (W^{\pm}\to\ell^{\pm}\nu)$, while the neutralino
decays to the lightest neutralino and the 125~GeV Higgs boson, $\ninotwo\to \ninoone (h\to bb/\gamma\gamma/\ell^{\pm}\nu qq)$.
The final states considered for the search have large missing
transverse momentum, an isolated electron or muon,
and one of the following: either
two jets identified as originating from bottom quarks, or two photons,
or a second electron or muon with the same electric charge.
The analysis is based on \lumi\ of $\sqrt{s}=8\TeV$ proton--proton collision data delivered by the Large Hadron Collider and recorded with the ATLAS detector.
Observations are consistent with the Standard Model expectations, and
limits are set in the context of a simplified supersymmetric model.
\end{sloppypar}
}

\AtlasTitle{\thetitle}
\AtlasAbstract{\theabstract}
\AtlasJournal{Eur. Phys. J. C}
\PreprintIdNumber{CERN-PH-EP-2014-298}

\begin{document}

%\linenumbers

\title{\thetitle}
\titlerunning{Search for charginos and neutralinos decaying to the Higgs boson}
\author{The ATLAS Collaboration}
\institute{CERN, 1211 Geneva 23, Switzerland, \email{atlas.publications@cern.ch}}
%\date{Received: \today / Revised version: \today}

\maketitle

\abstract{\theabstract}

%%%%%%%%%%%%%%%%%%%%%%%%%%%%%%%%%%%%%%%%%%%%%%%%%%
\section{Introduction}

\begin{sloppypar}
Supersymmetry
(SUSY)~\cite{Miyazawa:1966,Ramond:1971gb,Golfand:1971iw,Neveu:1971rx,Neveu:1971iv,Gervais:1971ji,Volkov:1973ix,Wess:1973kz,Wess:1974tw}
proposes the existence of new particles with spin
differing by one half unit from that of their Standard
Model (SM) partners.
In the Minimal Supersymmetric Standard Model (MSSM)~\cite{Fayet:1976et,Fayet:1977yc,Farrar:1978xj,Fayet:1979sa,Dimopoulos:1981zb},
charginos, $\chinopmonetwo$, and neutralinos, $\ninoonetwothreefour$, 
are the mass-ordered eigenstates formed from the linear superposition of the SUSY partners of the Higgs and electroweak gauge bosons (higgsinos, winos and bino).
In $R$-parity-conserving models, SUSY particles are pair-produced in colliders and the lightest SUSY particle (LSP) is stable. In many models the LSP is assumed to be a bino-like $\ninoone$, which is weakly interacting. 
Naturalness arguments~\cite{Barbieri:1987fn,deCarlos:1993yy} suggest that the lightest of the charginos and neutralinos may have masses at the electroweak scale, and may be accessible at the Large Hadron Collider (LHC)~\cite{Evans:2008zzb}.
Furthermore, direct pair production of charginos and neutralinos may be the dominant production of supersymmetric particles if the superpartners of the gluon and quarks are heavier than a few TeV.
\end{sloppypar}

In SUSY scenarios where the masses of the pseudoscalar Higgs boson and the superpartners of the leptons are larger than those of the produced chargino and neutralino, the chargino decays to the lightest neutralino and the $W$ boson, while the next-to-lightest neutralino decays to the lightest neutralino and the SM-like Higgs or $Z$ boson. This paper focuses on SUSY scenarios where the decay to the Higgs boson is the dominant one. This happens when the mass splitting between the two lightest neutralinos is larger than the Higgs boson mass and the higgsinos are much heavier than the winos, causing the composition of the lightest chargino and next-to-lightest neutralino to be wino-like and nearly mass degenerate. 

A simplified SUSY model~\cite{Alwall:2008ag,ArkaniHamed:2007fw} is considered for the optimisation of the search and the interpretation of results. 
It describes the direct production of $\chinoonepm$ and $\ninotwo$, where the masses and the decay modes of the relevant particles ($\chinoonepm$, $\ninoone$, $\ninotwo$) are the only free parameters. 
It is assumed that the $\chinoonepm$ and $\ninotwo$ are pure wino states and degenerate in mass, while the $\ninoone$ is a pure bino state.
The prompt decays $\chinoonepm\to W^\pm\ninoone$ and $\ninotwo\to h\ninoone$ are assumed to have 100\% branching fractions.
The Higgs boson mass is set to $125\GeV$, which is consistent with the measured value~\cite{Aad:2014aba}, and its branching fractions are assumed to be the same as in the SM. The latter assumption is motivated by those SUSY models in which the mass of the pseudoscalar Higgs boson is much larger than the $Z$ boson mass.

\begin{figure*}
\centering\hfill
\subfigure[One lepton and two $b$-quarks channel]{\includegraphics{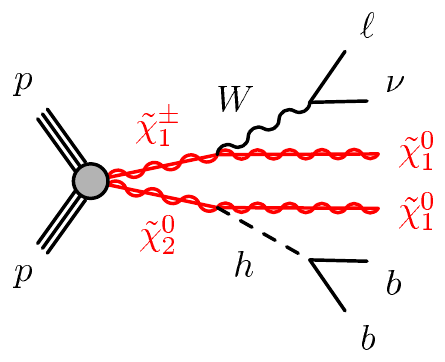}}\hfill\hfill
\subfigure[One lepton and two photons channel]{\includegraphics{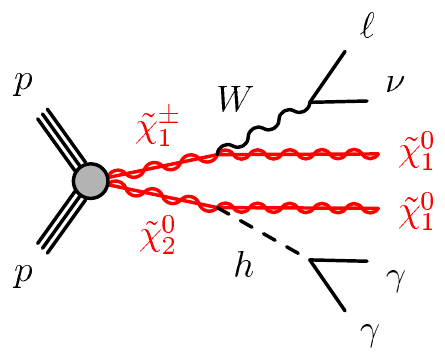}}\hfill\hfill
\subfigure[Same-sign dilepton channel]{\includegraphics{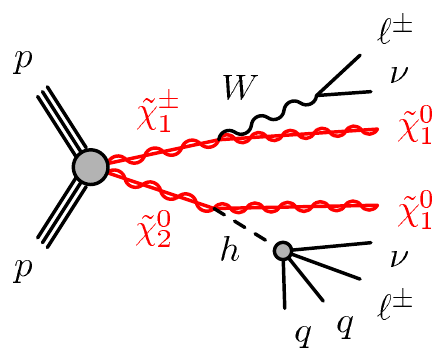}}\hfill\ 
\caption{
Diagrams for the direct pair production of
$\chinoonepm\ninotwo$ and the three decay modes studied in this paper.
For the same-sign dilepton channel (c), only the dominant decay mode is shown.
\label{fig:Feyn}}
\end{figure*}

The search presented in this paper targets leptonic decays of the $W$ boson and three Higgs boson decay modes as illustrated in Fig.~\ref{fig:Feyn}. The Higgs boson decays into a pair of $b$-quarks, or a pair of photons, or a pair of $W$ bosons where at least one of the bosons decays leptonically. The final states therefore contain missing transverse momentum from neutrinos and neutralinos, one lepton ($\ell=e$ or $\mu$), and one of the following: two $b$-quarks (\lbb), or  two photons (\lgg), or an additional lepton with the same electric charge (\llss).
The Higgs boson candidate can be fully reconstructed with the \lbb{}
and \lgg{} signatures. The \llss{} signature does not allow for such
reconstruction and it is considered because of its small SM
background. Its main
signal contribution is due to $h\to{}WW$, with smaller contributions
from $h\to{}ZZ$ and $h\to{}\tau\tau$ when some of the visible decay
products are missed during the event reconstruction. 

The analysis is based on \lumi\ of $\sqrt{s}=8\TeV$ proton--proton collision data delivered by the LHC and recorded with the ATLAS detector.
Previous searches for charginos and neutralinos at the LHC have been reported by the
ATLAS~\cite{ATLAS:3L8TeV,ATLAS:2L8TeV,Aad:2014yka} and
CMS~\cite{Khachatryan:2014qwa,Khachatryan:2014mma} collaborations.  Similar searches were conducted at the
Tevatron~\cite{D0-2009,CDF-2008} and
LEP~\cite{LEPSUSYWG:01-03.1,Heister:2003zk,Abdallah:2003xe,Acciarri:1999km,Abbiendi:2003sc}.

The results of this paper are combined with those of the ATLAS search using the three-lepton and missing transverse momentum final state, performed with the same dataset~\cite{ATLAS:3L8TeV}.
The three-lepton selections may contain up to two hadronically decaying $\tau{}$ leptons, providing sensitivity to the $h\rightarrow\tau\tau/WW/ZZ$ Higgs boson decay modes.
The statistical combination of the results is facilitated by the fact
that all event selections were constructed not to
overlap.

This paper is organised in the following way:
the ATLAS detector is briefly described in
Sect.~\ref{sec:atlas-detector}, followed by a description of the
Monte Carlo simulation in Sect.~\ref{sec:mc-simulation}. In
Sect.~\ref{sec:event-reco} the common aspects of the event
reconstruction are illustrated; Sects.~\ref{sec:lbb}, \ref{sec:lgg},
and \ref{sec:ss-2l} describe the channel-specific features;
Sect.~\ref{sec:systematics} discusses the systematic uncertainties;
the results and conclusions are presented in Sects.~\ref{sec:results}
and \ref{sec:conclusions}.

%%%%%%%%%%%%%%%%%%%%%%%%%%%%%%%%%%%%%%%%%%%%%%%

\section{The ATLAS detector}
\label{sec:atlas-detector}

\begin{sloppypar}
ATLAS is a multipurpose particle physics
experiment~\cite{atlas-det}. It consists of detectors forming a
forward-backward symmetric cylindrical geometry.\footnote{%
	ATLAS uses a right-handed coordinate
	system with its origin at the nominal interaction point (IP) in the
	centre of the detector and the $z$-axis along the beam line. The
	$x$-axis points from the IP to the centre of the LHC ring, and the
	$y$-axis points upward. Cylindrical coordinates $(r,\phi)$ are used
	in the transverse plane, $\phi$ being the azimuthal angle around the
	$z$-axis. The pseudorapidity is defined in terms of the polar angle
	$\theta$ as $\eta=-\ln\tan(\theta/2)$.}
The inner detector (ID) covers $|\eta|\,$$<\,$2.5 and consists of a
silicon pixel detector, a semiconductor microstrip tracker, and a
transition radiation tracker.
The ID is surrounded by a thin superconducting solenoid providing a 2$\,$T axial magnetic field.
A high-granularity lead/liquid-argon (LAr) sampling calorimeter measures the energy and the position of electromagnetic showers within $|\eta|\,$$<\,$3.2.
Sampling calorimeters with LAr are also used to measure hadronic showers in the endcap (1.5$\,$$<\,$$|\eta|\,$$<\,$3.2) and forward (3.1$\,$$<\,$$|\eta|\,$$<\,$4.9) regions, while a steel/scintil\-la\-tor tile calorimeter measures hadronic showers in the central region ($|\eta|\,$$<\,$1.7).
The muon spectrometer (MS) surrounds the calorimeters and consists of three large superconducting air-core toroid magnets, each with eight coils, precision tracking chambers ($|\eta|\,$$<\,$2.7), and fast trigger chambers ($|\eta|\,$$<\,$2.4).
A three-level trigger system selects events to be recorded for permanent storage.
\end{sloppypar} 
 
%%%%%%%%%%%%%%%%%%%%%%%%%%%%%%%%%%%%%%%%%%%%%%%

\section{Monte Carlo simulation}
\label{sec:mc-simulation}

\begin{table*}
\centering
\caption{\label{tab:MCsamples}
	Simulated samples used for background estimates.
	``Tune'' refers to the choice of parameters used for the underlying-event generation.
}
\scalebox{0.94}{
\footnotesize
\begin{tabular}{lllll}
\toprule
Process & Generator & Cross section  & Tune   & PDF set  \\
\midrule
Single top, $t$-channel & \acermc{}~\cite{Kersevan:2004yg}+\pythia~\cite{Sjostrand:2006za}    & NNLO+NNLL~\cite{Kidonakis:2011wy}  & \auet~\cite{ATLAS:2011zja} & \cteqsixl~\cite{Pumplin:2002vw} \\
Single top, $s$-channel & \powheg~\cite{Nason:2004rx,Frixione:2007vw}+\pythia    & NNLO+NNLL~\cite{Kidonakis:2010tc} & \perugia~\cite{Skands:2010ak} &  \ctten~\cite{CT10pdf} \\
$tW$ & \powhegpythia   & NNLO+NNLL~\cite{Kidonakis:2010ux} & \perugia & \ctten \\
$\ttbar$ & \powhegpythia    & NNLO+NNLL~\cite{Cacciari:2011hy,Baernreuther:2012ws,Czakon:2012zr,Czakon:2012pz,Czakon:2013goa,Czakon:2011xx} & \perugia & \ctten \\ 
$\ttbar W$, $\ttbar Z$ & \madgraph~\cite{Alwall:2007st}+\pythia &  NLO & \auet & \cteqsixl \\ 
$W$, $Z$ (\lbb{} channel)    & \sherpa~\cite{Sherpa} & NLO & -- & \ctten \\ 
$W$, $Z$ (\llss{} channel) & \alpgen{}~\cite{Mangano:2002ea}+\pythia & NLO & \perugia & \cteqsixl \\ 
$WW$, $WZ$, $ZZ$ & \sherpa & NLO & -- & \ctten \\
$W\gamma$ $W\gamma\gamma$  & \alpgen{}+\pythia & NLO & \auet & \cteqsixl \\
$Z\gamma$, $Z\gamma\gamma$ & \sherpa & NLO & -- & \ctten \\
$Wh$, $Zh$ & \pythiaeight~\cite{Sjostrand:2007gs} & NNLO(QCD)+NLO(EW)~\cite{CERNYellowReport3} & \au~\cite{ATLAS:2012uec} & \cteqsixl \\
$\tth$ & \pythiaeight & NLO(QCD)~\cite{CERNYellowReport3} & \au & \cteqsixl \\
\bottomrule
\end{tabular}}
\end{table*}

The event generators, the accuracy of theoretical cross sections, the underlying-event parameter
tunes, and the parton distribution function (PDF) sets used for simulating the
SM background processes are summarised in Table~\ref{tab:MCsamples}.

The SUSY signal samples  are produced with \herwigpp~\cite{herwigplusplus} using the \cteqsixl\ PDF set. 
Signal cross sections are calculated at next-to-leading order (NLO) 
in the strong coupling constant using \prospino~\cite{Beenakker:1996ch}.
These agree with the NLO calculations matched to resummation at next-to-leading-logarithmic (NLL) accuracy within $\sim$2\%~\cite{Fuks:2012qx,Fuks:2013vua}.
For each cross section, the nominal value and its uncertainty are taken respectively from the centre and the spread of the cross-section predictions using different PDF sets and their associated uncertainties, as well as from variations of factorisation and renormalisation scales, as described in Ref.~\cite{Kramer:2012bx}.

The propagation of
particles through the ATLAS detector is modelled with
\geant~\cite{Agostinelli:2002hh} using the full ATLAS detector
simulation~\cite{:2010wqa} for all Monte Carlo (MC) simulated samples, except for $\ttbar$ production and the SUSY signal samples in which the Higgs boson decays to two $b$-quarks, for which a fast simulation based on a parametric response of
the electromagnetic and hadronic calorimeters is used~\cite{atlfastII}. 
The effect of multiple proton--proton collisions in the same or nearby beam bunch crossings (in-time or out-of-time pile-up) is incorporated into the simulation by overlaying additional minimum-bias events generated with \pythia{} onto hard-scatter events. 
Simulated events are weighted so that the distribution of the average number of interactions per bunch crossing matches that observed in data, but are otherwise reconstructed in the same manner as data. 

%%%%%%%%%%%%%%%%%%%%%%%%%%%%%%%%%%%%%%%%%%%%%%%

\section{Event reconstruction}
\label{sec:event-reco}

The data sample considered in this analysis was collected with a
combination of single-lepton, dilepton, and diphoton triggers. After
applying beam, detector, and data-quality requirements, the dataset
corresponds to an integrated luminosity of \lumi{}, with an
uncertainty of \lumierror{} derived following the
methodology detailed in Ref.~\cite{Aad:2013ucp}.

Vertices compatible with the proton-proton interactions are reconstructed using tracks from the ID.
Events are analysed if the primary vertex has five or more tracks,
each with transverse momentum $\pt>400\MeV$, unless stated otherwise.
The primary vertex of an event is identified as the vertex with the largest $\sum\pt^2$ of the associated tracks.

\begin{sloppypar}
Electron candidates are reconstructed from calibrated clustered energy deposits in the electromagnetic calorimeter and a matched ID track, which in turn determine the $\pt$ and $\eta$ of the candidates respectively.
Electrons must satisfy ``medium'' cut-based identification criteria, following Ref.~\cite{Aad:2014fxa}, and are required to have $\pt>10\GeV$ and $|\eta|<2.47$.
\end{sloppypar}

Muon candidates are reconstructed by combining tracks in the ID and tracks or segments in the MS~\cite{Aad:2014rra} and are required to have $\pt>10\GeV$ and $|\eta|<2.5$. 
To suppress cosmic-ray muon background, events are rejected if they
contain a muon having transverse impact parameter with respect to the
primary vertex $|d_0|>0.2$\,mm or longitudinal impact parameter with
respect to the primary vertex $|z_0|>1$\,mm.

Photon candidates are reconstructed from clusters of energy deposits in the electromagnetic calorimeter.
Clusters without matching tracks as well as those matching one or two tracks consistent with a photon
conversion are considered.
The shape of the cluster must match that expected for an electromagnetic shower, using criteria 
tuned for robustness under the pile-up conditions of 2012~\cite{Aad:2014nim}.
The cluster energy is calibrated separately for converted and unconverted photon candidates using simulation.
In addition, $\eta$-dependent correction factors determined from $Z\to e^+e^-$ events are applied to the cluster energy, as described in Ref.~\cite{Aad:2014nim}.
The photon candidates must have $\pt>20\GeV$ and $|\eta|<2.37$, excluding the transition region $1.37<|\eta|<1.56$
between the central and endcap electromagnetic calorimeters.  The tighter $\eta$ requirement on photons, as compared to electrons,
 reflects the poorer photon resolution in the transition region and for $2.37\leq|\eta|<2.47$.

Jets are reconstructed with the anti-$k_t$ algorithm~\cite{Cacciari:2008gp} with a radius parameter of 0.4 using three-dimensional clusters of energy in the calorimeter~\cite{Lampl:2008zz} as input. 
The clusters are calibrated, weighting differently the energy deposits arising from the electromagnetic and hadronic components of the showers.
The final jet energy calibration corrects the calorimeter response to the particle-level jet energy~\cite{Aad:2011he,Aad:2014bia}; the correction factors are obtained from simulation and then refined and validated using data.
Corrections for in-time and out-of-time pile-up are also applied, as described in Ref.~\cite{TheATLAScollaboration:2013pia}.
Events containing jets failing to meet the quality criteria described in Ref.~\cite{Aad:2011he} are rejected to suppress
non-collision background and events with large noise in the calorimeters.

Jets with $\pt>20\GeV$ are considered in the central pseudorapidity ($|\eta|<2.4$) region, and jet $\pt>30\GeV$ is required in the forward ($2.4<|\eta|<4.5$) region.
For central jets, the \pt\ threshold is lower 
since it is possible to suppress pile-up using information from the ID, the ``jet vertex fraction'' (JVF).
This is defined as the \pt-weighted fraction of tracks within the jet that originate from the primary vertex of the event,
and is $-1$ if there are no tracks within the jet.
Central jets can also be tagged as originating from bottom quarks (referred to as $b$-jets)
using the MV1 multivariate $b$-tagging algorithm based on quantities related to impact parameters of tracks and reconstructed secondary vertices~\cite{btag}. 
The efficiency of the $b$-tagging algorithm depends on the operating
point chosen for each channel, and is reported in
Sects.~\ref{sec:lbb} and \ref{sec:ss-2l}.

Hadronically decaying $\tau$ leptons are 
reconstructed as 1- or 3-prong hadronic jets within $|\eta|<2.47$,
and are required to have $\pt>20\GeV$ after being calibrated to the $\tau$ energy scale~\cite{Aad:2014rga}.
Final states with hadronically  decaying $\tau$ leptons are not considered here; however, identified $\tau$ leptons are used in the overlap removal procedure described
below, as well as to ensure that the same-sign lepton channel does not overlap with
the three-lepton search~\cite{ATLAS:3L8TeV} that is included in the
combined result.

\begin{sloppypar}
Potential ambiguities between candidate leptons, photons and jets
are resolved by removing one or both objects if they are separated by
$\Delta{}R\equiv\sqrt{(\Delta\phi)^2+(\Delta\eta)^2}$ below a threshold. This process eliminates duplicate objects
reconstructed from a single particle, and suppresses leptons and
photons contained inside hadronic jets. The thresholds and the order in
which overlapping objects are removed are summarised in
Table~\ref{tab:overlap}.
In the same-sign channel, $e^+e^-$ and
$\mu^+\mu^-$ pairs with $m_{\ell^+\ell^-}<12~\GeV{}$ are also removed.
The remaining leptons and photons are referred to as ``preselected" objects.
\end{sloppypar}

\begin{table}
\centering
\caption{\label{tab:overlap}
Summary of the overlap removal procedure. Potential ambiguities are
resolved by removing nearby objects in the indicated order, from top
to bottom. Different $\Delta{}R$ separation requirements are used in
the three channels. 
}
\begin{tabular}{llllr}
\toprule
Candidates &  \multicolumn{3}{c}{$\Delta{}R$ threshold} & Candidate removed \\
\cmidrule{2-4}
                             &  \lbb{} & \lgg{} & \llss{}  & \\
\midrule
$e$--$e$                        & 0.1     & ---    & 0.05     & lowest-\pt{} $e$  \\
$e$--$\gamma$                   & ---     & 0.4    & ---      & $e$               \\
jet--$\gamma$                 & ---     & 0.4    & ---      & jet               \\
jet--$e$                      & 0.2     & 0.2    & 0.2      & jet               \\
$\tau{}$--$e$ or $\tau{}$--$\mu{}$ & ---     & ---    & 0.2      & $\tau$            \\
$\mu$--$\gamma$                 & ---     & 0.4    & ---      & $\mu$             \\
$e$--jet or $\mu$--jet         & 0.4     & 0.4    & 0.4      & $e$ or $\mu$      \\
$e$--$\mu$                      & $0.1$   & ---    & $0.1$    & both              \\
$\mu$--$\mu$                    & $0.05$  & ---    & $0.05$   & both              \\
jet--$\tau$                   & ---     & ---    & 0.2      & jet               \\
\bottomrule
\end{tabular}
\end{table} 

\begin{sloppypar}
Isolation criteria are applied to improve the purity of reconstructed objects.
The criteria are based on the scalar sum of the transverse energies \et{} of the calorimeter cell clusters within a radius $\Delta{}R$ of the object (\etcone{\Delta{}R}), and on the scalar sum of the \pt{} of the tracks within $\Delta{}R$ and associated with the primary vertex (\ptcone{\Delta{}R}).
The contribution due to the object itself is not included in either sum.
The values used in the isolation criteria depend on the channel;
they are specified in Sects.~\ref{sec:lbb}, \ref{sec:lgg} and
\ref{sec:ss-2l}.
\end{sloppypar}

The missing transverse momentum, $\pTmiss$ (with magnitude $\met$), is the negative vector sum of the transverse momenta of all preselected electrons, muons, and photons, as well as jets and calorimeter energy clusters with $|\eta|\,$$<\,$4.9 not associated with these objects. 
Clusters that are associated with electrons, photons and jets are calibrated to
the scale of the corresponding objects~\cite{Aad:2012re,Aad:2013oia}.

The efficiencies for electrons, muons, and photons to satisfy the
reconstruction and identification criteria are measured in control
samples, and corrections are applied to the simulated samples to
reproduce the efficiencies in data. Similar corrections are also
applied to the trigger efficiencies, as well as to the jet $b$-tagging
efficiency and misidentification probability.

%%%%%%%%%%%%%%%%%%%%%%%%%%%%%%%%%%%%%%%%%%%%%%%

\section{One lepton and two $b$-jets channel}
\label{sec:lbb}

\begin{table*}[t]\centering
\caption{\label{tab:SRlbb}
Selection requirements for the signal, control and validation regions of the one lepton and two $b$-jets channel.
The number of leptons, jets, and $b$-jets is labelled with \nlep, \njet, and \nbjet respectively.}
\begin{tabular}{ccccccc}
\toprule
& \SRlbbone & \SRlbbtwo & \CRlbbT & \CRlbbW &  \VRlbbone & \VRlbbtwo \\
\midrule
\nlep\ & 1 & 1 & 1 & 1 & 1 & 1 \\
\njet\ & 2--3 & 2--3 & 2--3 & 2 & 2--3 & 2--3 \\
\nbjet\ & 2 & 2 & 2 & 1 & 2 & 2\\
\met\ [GeV] & $>100$   & $>100$ & $>100$  & $>100$ & $>100$  & $>100$ \\
\mct\ [GeV] & $>160$   & $>160$ & 100--160 & $>160$ & 100--160 & $>160$ \\
\mtW\ [GeV] & 100--130 & $>130$ & $>100$  & $>40$ & 40--100 & 40--100 \\
\bottomrule
\end{tabular}
\end{table*}

%%%%%%%%%%%%%%%%%%%%%%%%%%%%%%%%%%%%%%%%%%%%%%%

\subsection{Event selection}

The events considered in the one lepton and two $b$-jets channel are recorded
with a combination of single-lepton triggers with a \pt\ threshold of 24 GeV.
To ensure that the event is triggered with a constant high efficiency,
the offline event selection requires exactly one signal lepton ($e$ or
$\mu$) with $\pt>25\GeV$.
The signal electrons must satisfy the ``tight'' identification criteria
of Ref.~\cite{Aad:2014fxa}, as well as $|d_0|/\sigma_{d_0}<5$, where
$\sigma_{d_0}$ is the error on $d_0$, and $|z_0\sin\theta|<0.4$\,mm.  The signal
muons must satisfy $|\eta|<2.4$, $|d_0|/\sigma_{d_0}<3$, and
$|z_0\sin\theta|<0.4$\,mm.  The signal electrons (muons) are required to satisfy
the isolation criteria $\etcone{0.3}/\pt<0.18$ (0.12) and
$\ptcone{0.3}/\pt<0.16$ (0.12).

Events with two or three jets are selected, and the jets can be
either central ($|\eta|<2.4$) or forward ($2.4<|\eta|<4.9$).  Central
jets have $\pt>25~\GeV$, and forward jets have $\pt>30~\GeV$.  For
central jets with $\pt<50\GeV$, the $\jvf{}$ must be $>0.5$.
Events must contain exactly two $b$-jets and these must be the highest-\pt{} central jets.
The chosen operating point of the $b$-tagging
algorithm identifies $b$-jets in simulated $\ttbar$ events
with an efficiency of 70\%; 
it misidentifies charm jets 20\% of the time and light-flavour (including gluon-induced)
jets less than $1\%$ of the time.  

After the requirement of $\met>100$~GeV, the dominant background contributions in the \lbb{} channel are \ttbar{},
\wjets{}, and single-top $Wt$ production. Their contributions are suppressed
using the kinematic selections described below, which define the
two signal regions (SR) \SRlbbone{} and \SRlbbtwo{} summarised in
Table~\ref{tab:SRlbb}.

The contransverse mass \mct~\cite{Tovey:2008ui,Polesello:2009rn}
is defined as
\begin{equation}
\mct = \sqrt{(E_\mathrm{T}^{b_1}+E_\mathrm{T}^{b_2})^2 - |\vec{p}_\mathrm{T}^{b_1}-\vec{p}_\mathrm{T}^{b_2}|^2},
\end{equation}
where $E_\mathrm{T}^{b_i}$ and $\vec{p}_\mathrm{T}^{b_i}$ are the transverse energy and momentum
of the $i$-th $b$-jet.
The SM \ttbar{} background has an upper endpoint at $\mct$ of approximately
$m_t$, and is efficiently suppressed by requiring $\mct>160\GeV$.

The transverse mass \mtW, describing $W$ candidates in background events, is defined as
\begin{equation}
\mtW = \sqrt{2 E_\mathrm{T}^\ell \met - 2 \vec{p}_\mathrm{T}^\ell \cdot \vec{p}_\mathrm{T}^\mathrm{miss}},
\end{equation}
where $E_\mathrm{T}^\ell$ and $\vec{p}_\mathrm{T}^\ell$ are the transverse energy and momentum of the lepton.
Requiring $\mtW>100\GeV$ efficiently suppresses the $W$\,+\,jets background.
The two SRs are distinguished by requiring $100<\mtW<130\GeV$ for \SRlbbone\ and $\mtW>130\GeV$ for \SRlbbtwo.
The first signal region provides sensitivity to signal models with a mass splitting between \ninoone\ and \ninotwo\  similar to the Higgs boson mass, while the second one targets larger mass splittings.

In each SR, events are classified into five bins of the invariant mass \mbb\ 
of the two $b$-jets
as 45--75--105--135--165--195\,GeV.
In the SRs, about 70\% of the signal events due to $h\to\bbbar$ populate the central bin of
105--135\,\GeV. The other four bins (sidebands) are used to
constrain the background normalisation, as described below.

\begin{figure*}\centering
\subfigure[\mct\ in \CRlbbT, \SRlbbone\ and \SRlbbtwo, central \mbb\ bin]
	{\includegraphics[width=0.39\textwidth]{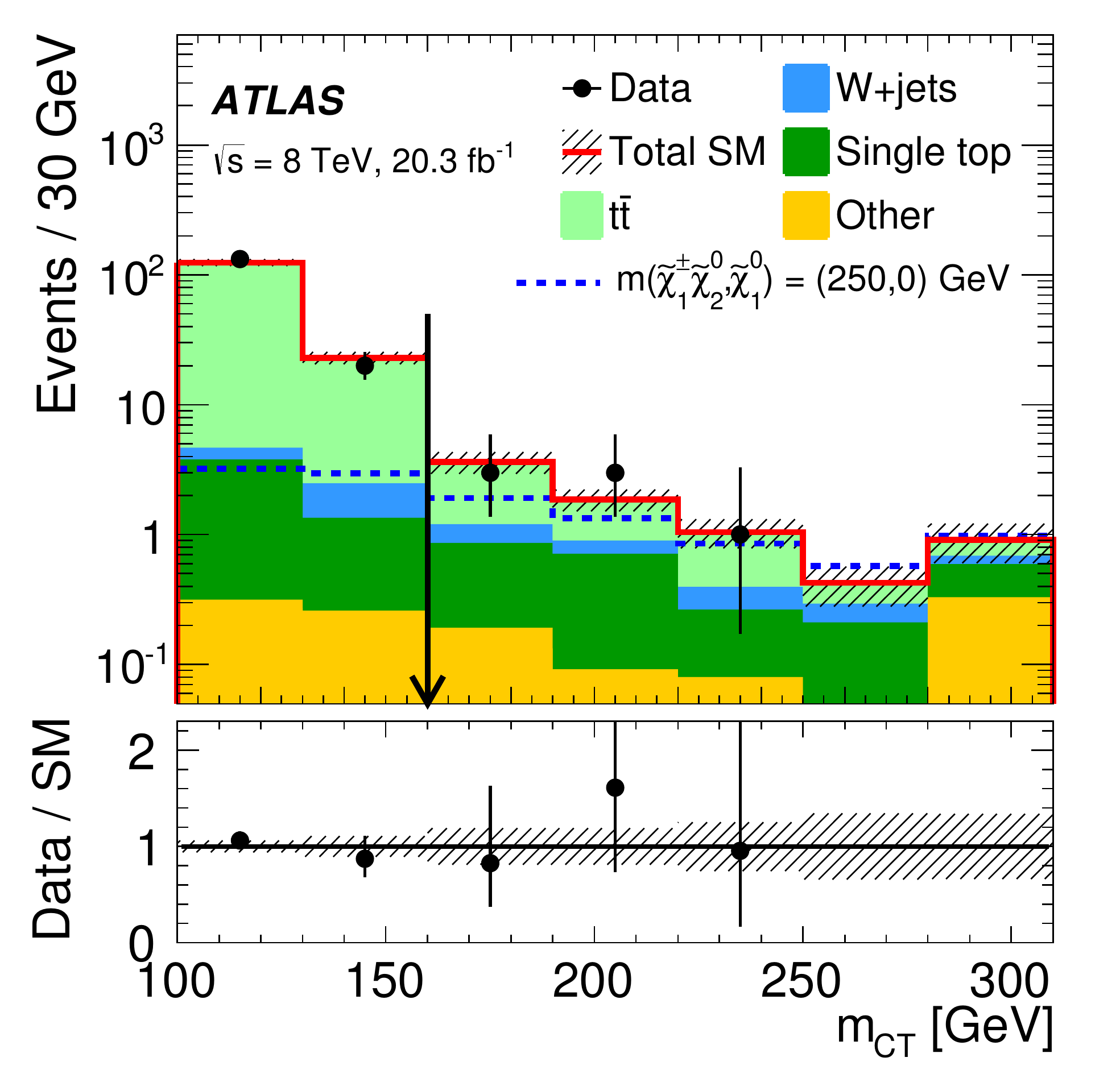}}\hspace{1.8cm}
\subfigure[\mct\ in \CRlbbT, \SRlbbone\ and \SRlbbtwo, \mbb\ sidebands]
	{\includegraphics[width=0.39\textwidth]{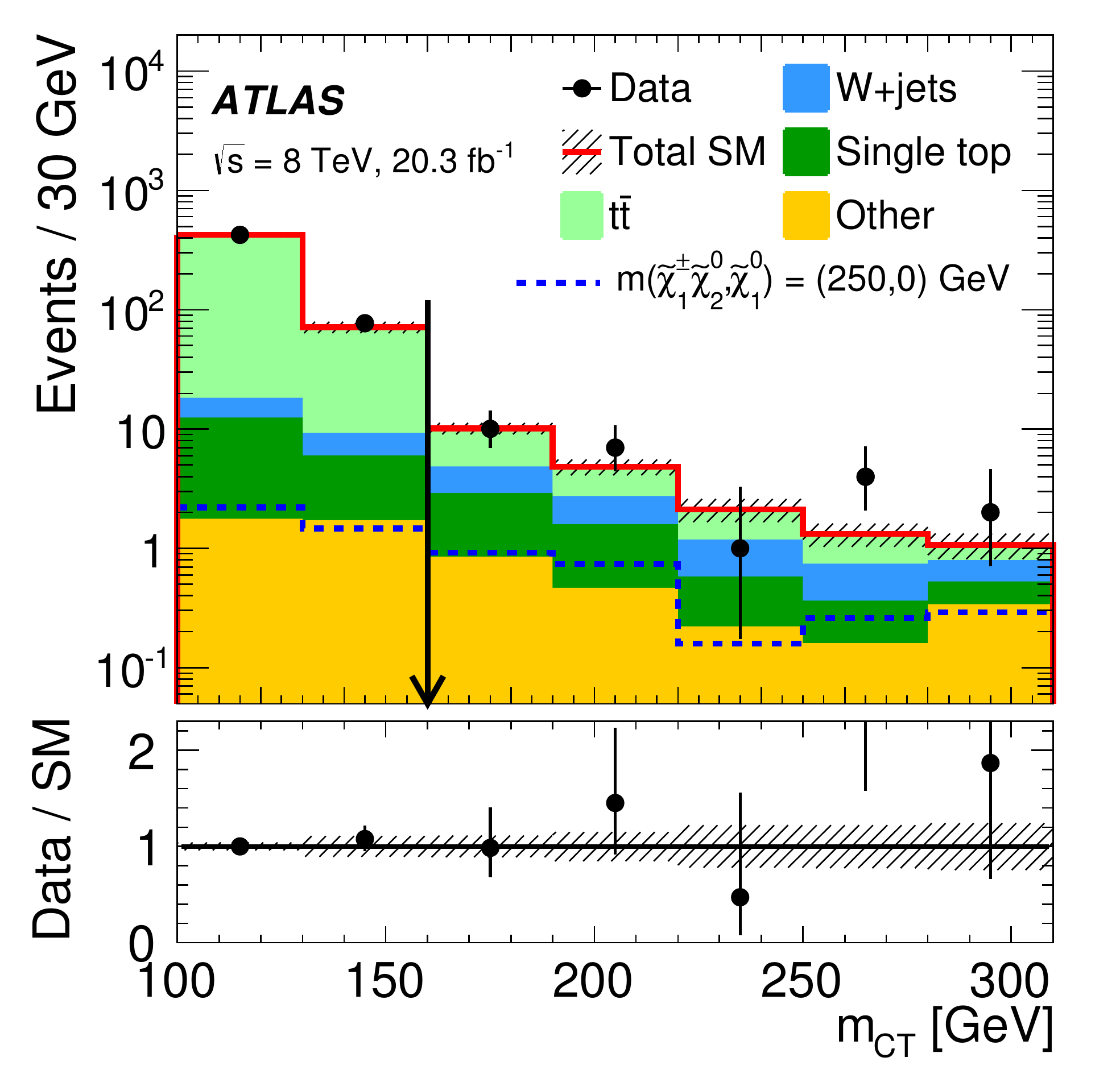}}
\subfigure[\mtW\ in \VRlbbtwo, \SRlbbone\ and \SRlbbtwo, central \mbb\ bin]
	{\includegraphics[width=0.39\textwidth]{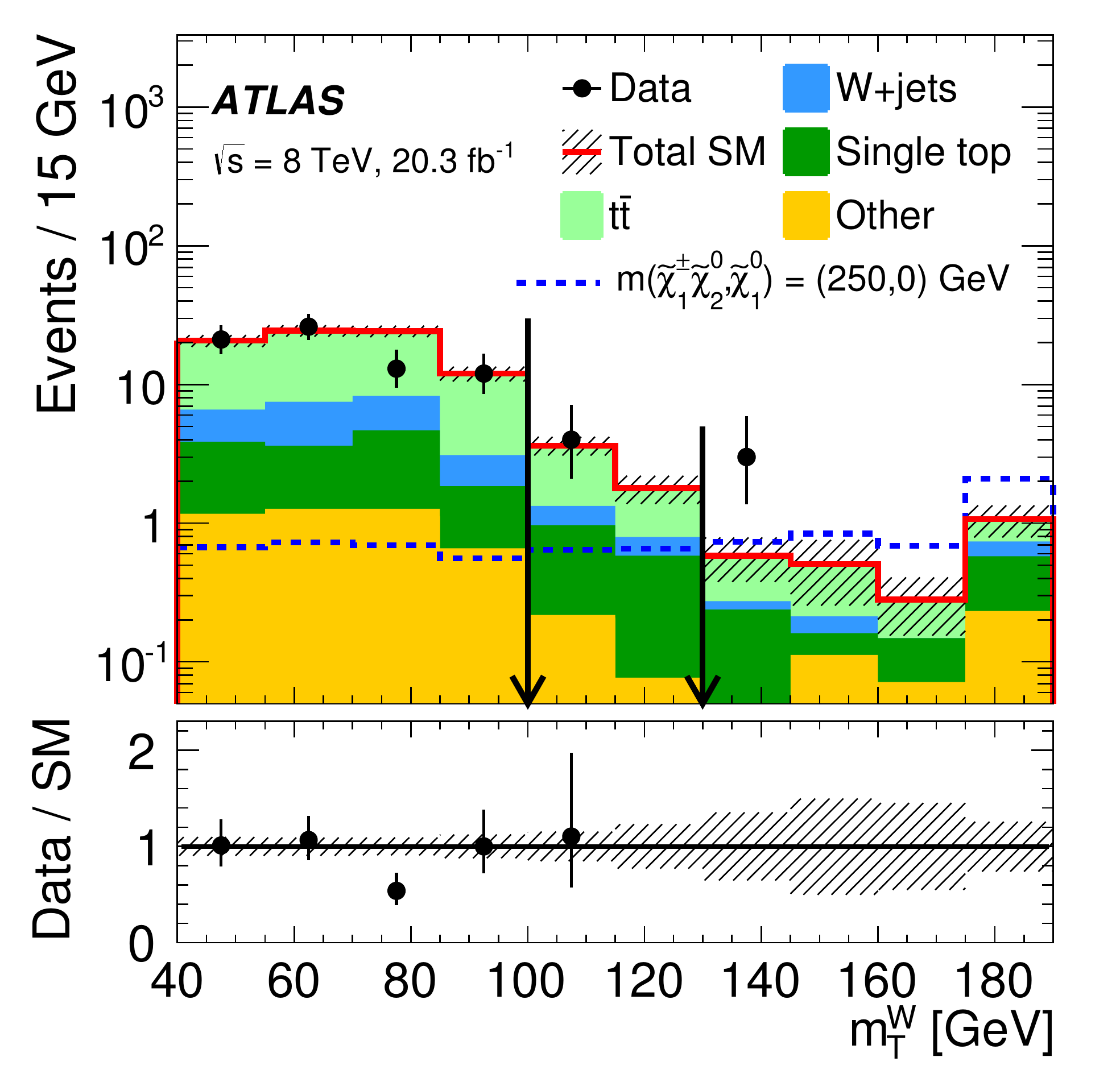}}\hspace{1.8cm}
\subfigure [\mtW\ in \VRlbbtwo, \SRlbbone\ and \SRlbbtwo, \mbb\ sidebands]
	{\includegraphics[width=0.39\textwidth]{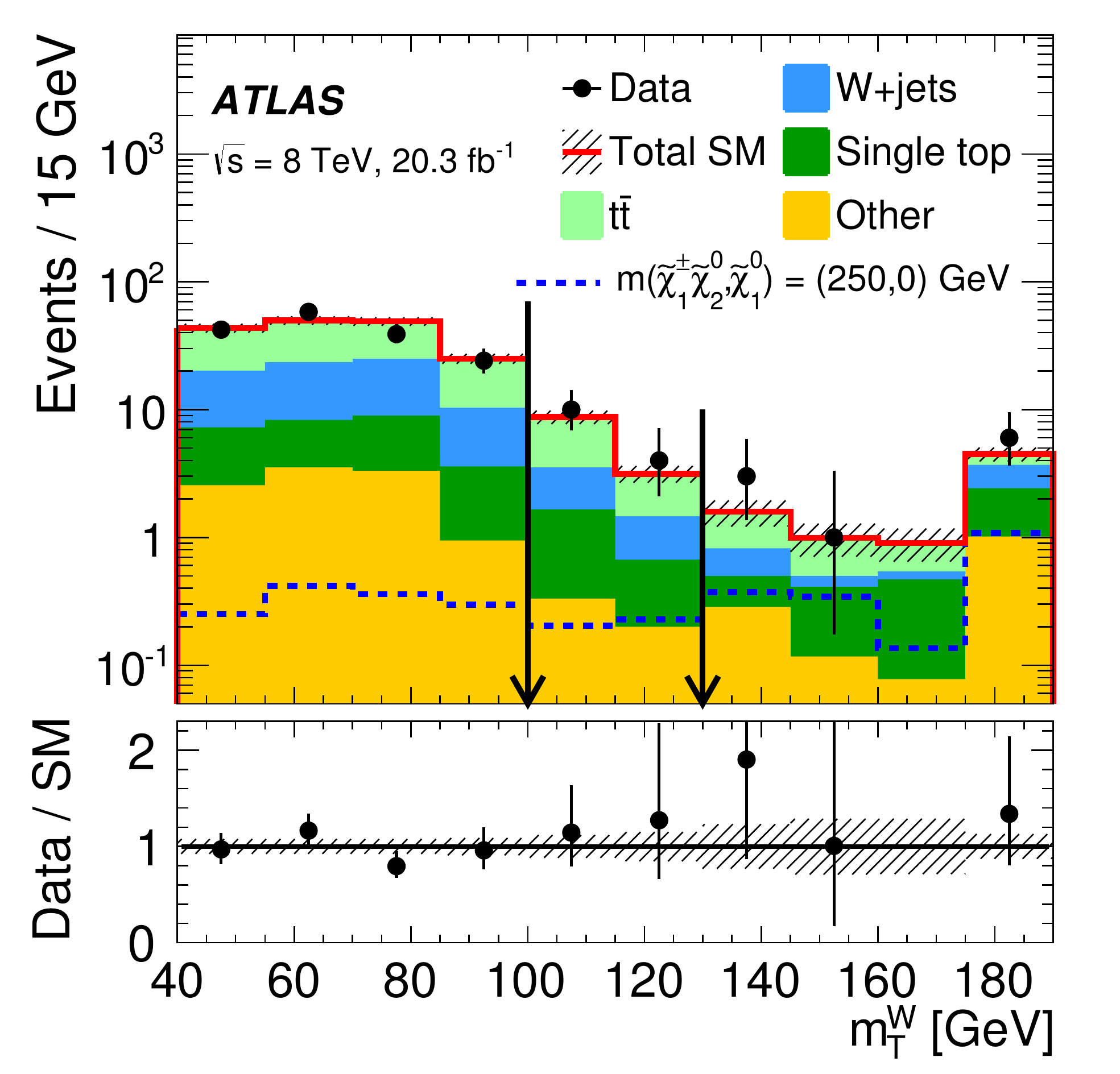}}
\subfigure[Number of $b$-jets in \SRlbbone\ and \SRlbbtwo\ without the $b$-jet multiplicity requirement, central \mbb\ bin]
	{\includegraphics[width=0.39\textwidth]{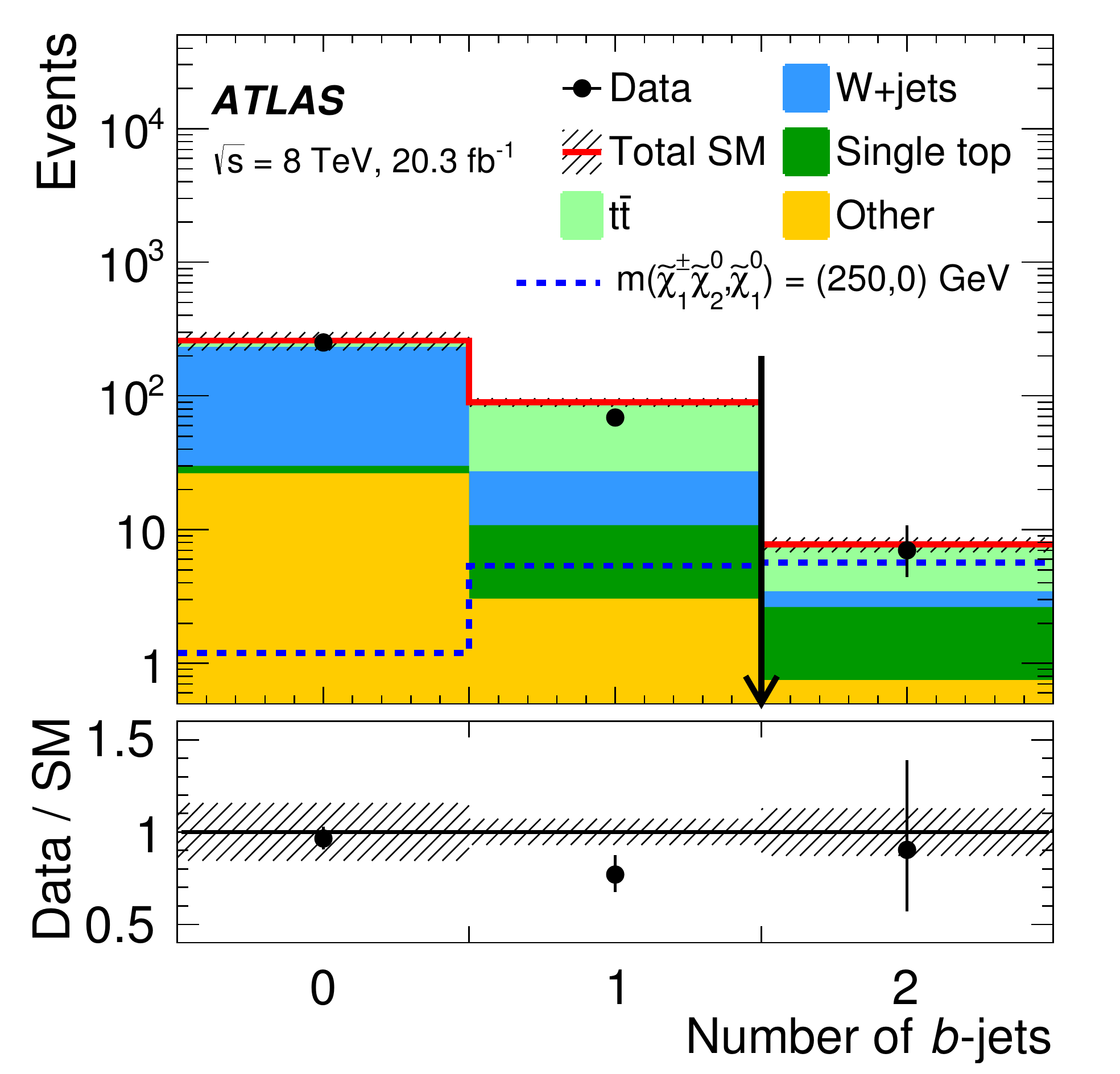}}\hspace{1.8cm}
\subfigure[\mbb\ in \SRlbbone\ and \SRlbbtwo]
	{\includegraphics[width=0.39\textwidth]{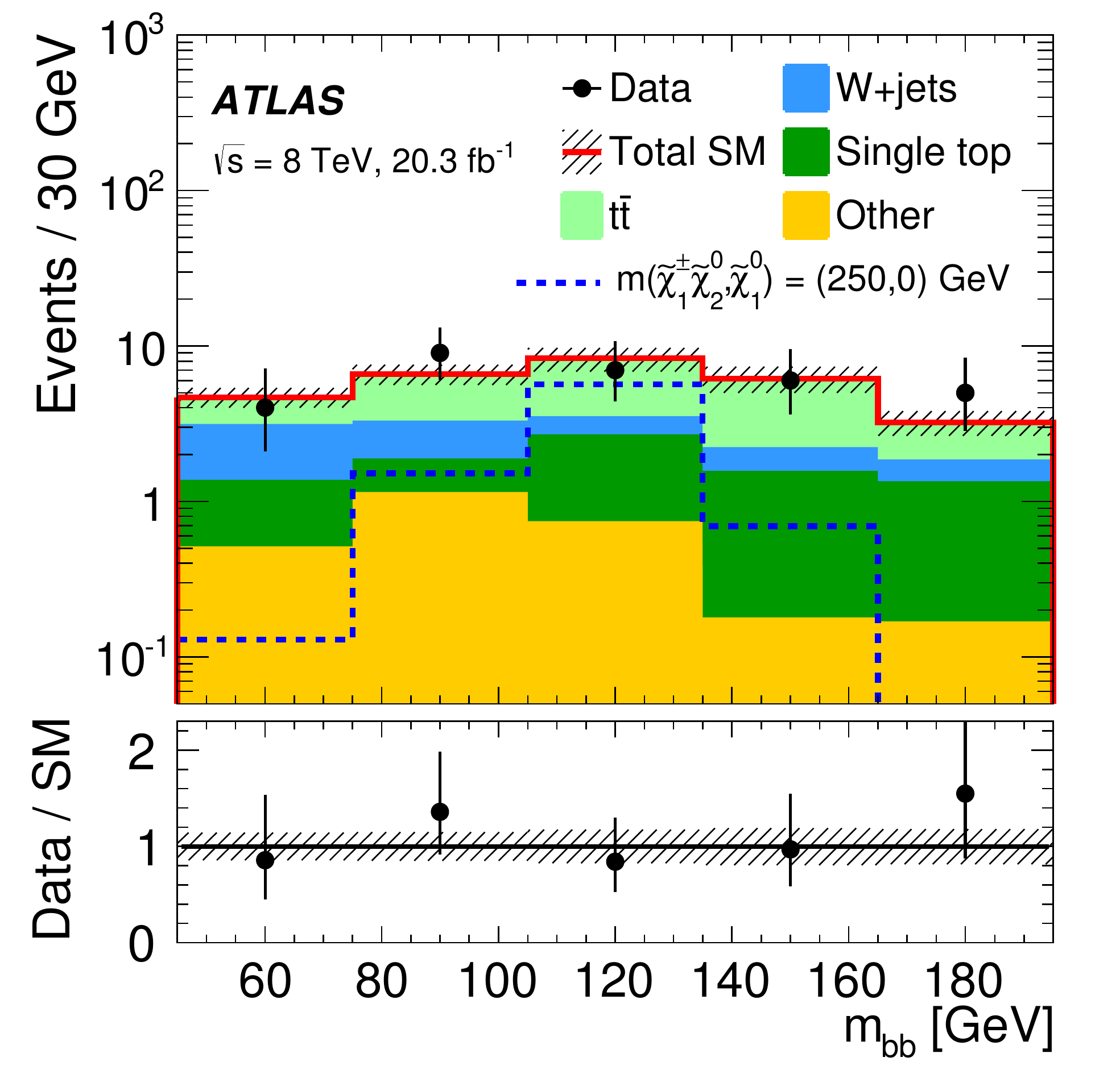}}
\caption{\label{fig:SRlbb}
Distributions of contransverse mass \mct, transverse mass of the $W$-candidate \mtW, number of $b$-jets, and invariant mass of the $b$-jets \mbb\ for the one lepton and two $b$-jets channel in the indicated regions.
The stacked background histograms are obtained from the background\hyp only fit.
The hashed areas represent the total uncertainties on the background estimates after the fit.
The rightmost bins in (a)--(d) include overflow.
The distributions of a signal hypothesis are also shown without stacking on the background histograms.
The vertical arrows indicate the boundaries of the signal regions.
The lower panels show the ratio of the data to the SM background prediction.
}
\end{figure*}

\begin{table*}[th!]
\caption{
Event yields and SM expectation in the one lepton and two $b$-jets channel
obtained with the background\hyp only fit.
``Other'' includes $Z$ + jets, $WW$, $WZ$, $ZZ$, $Zh$ and $Wh$ processes.
The errors shown include statistical and systematic uncertainties.
\label{tab:results_blinded_bkgOnlyFit}}
\begin{center}
\footnotesize
\begin{tabular}{l*{8}{r@{\,$\pm$\,}r}}
\toprule
& \multicolumn{2}{c}{\SRlbbone}
& \multicolumn{2}{c}{\ \ \ \SRlbbtwo}
& \multicolumn{2}{c}{\SRlbbone}
& \multicolumn{2}{c}{\SRlbbtwo}
& \multicolumn{2}{c}{\CRlbbT}
& \multicolumn{2}{c}{\CRlbbW}
& \multicolumn{2}{c}{\VRlbbone}
& \multicolumn{2}{c}{\VRlbbtwo}\\
& \multicolumn{4}{c}{$105<\mbb<135$ \GeV}
& \multicolumn{4}{c}{\mbb\ sidebands}\\
\midrule
Observed events & \multicolumn{2}{l}{4}   & \multicolumn{2}{l}{\ \ \ 3}
                & \multicolumn{2}{l}{14}  & \multicolumn{2}{l}{10}
                & \multicolumn{2}{l}{651} & \multicolumn{2}{l}{1547}
                & \multicolumn{2}{l}{885} & \multicolumn{2}{l}{235} \\
SM expectation  & 6.0&1.3 & \ \ \ 2.8&0.8 & 13.1&2.4 & \phantom{0}8.8&1.7 & 642&25 & 1560&40 & 880&90 & 245&17 \\
\midrule
\ttbar\         & 3.8&1.2 & \ \ \ 1.4&0.7 & 8.0&2.4 & 3.1&1.4 & 607&25 & 680&60 & 680&90 & 141&18 \\
$W$ + jets      & 0.6&0.3 & \ \ \ 0.2&0.1 & 2.7&0.5 & 1.7&0.3 &  11& 2 & 690&60 &  99&12 &  62& 8 \\
Single top      & 1.3&0.4 & \ \ \ 0.7&0.4 & 1.9&0.6 & 2.5&1.1 &  20& 4 & 111&14 &  80&10 &  27& 4 \\
Other & 0.3&0.1 & \ \ \ 0.5&0.1 & 0.5&0.1 & 1.5&0.2 &   4& 1 &  76& 8 &  16& 2 &  15& 1 \\
\bottomrule
\end{tabular}
\end{center}
\end{table*}

%%%%%%%%%%%%%%%%%%%%%%%%%%%%%%%%%%%%%%%%%%%%%%%

\subsection{Background estimation}
\label{sec:lbb-bkg}

The contributions from the \ttbar{} and \wjets{} background sources are estimated from simulation, and normalised to data in dedicated control regions defined in the following paragraphs.
The contribution from multi-jet production, where the signal lepton is a misidentified jet or comes from a heavy-flavour hadron decay or photon conversion, is
estimated using the ``matrix method'' described in Ref.~\cite{ATLAS:2L8TeV},
and is found to be less than 3\% of the
total background in all regions and is thus neglected.
The remaining sources of background (single top, $Z$ + jets, $WW$, $WZ$, $ZZ$, $Zh$ and $Wh$ production) are estimated from simulation.

Two control regions (CR), \CRlbbT{} and \CRlbbW{}, are designed to
constrain the normalisations of the \ttbar{} and \wjets{} backgrounds respectively.
The acceptance for \ttbar{} events is increased in \CRlbbT{} by modifying the requirement
on \mct{} to $100<\mct<160\GeV$.  The acceptance of \wjets{}
events is increased in \CRlbbW{} by requiring $\mtW{}>40\GeV$ and
exactly two jets, of which only one is $b$-tagged. These two
control regions are summarised in Table~\ref{tab:SRlbb}.
The control regions are defined to be similar to the signal regions in order to reduce systematic uncertainties on the extrapolation to the signal regions; at the same time they are dominated by the targeted background processes and the expected contamination by signal is small.

As in the signal regions, the control regions are binned in
\mbb (\mbj in the case of \CRlbbW).
A ``background\hyp only" likelihood fit is performed,
in which the predictions of the simulated background processes without any signal hypothesis are fit simultaneously to the data yields in
eight \mbb{} sideband bins of the SRs and the ten \mbb{} bins of the CRs.
This fit, as well as the limit-setting procedure, is performed using
the \histfitter{} package described in Ref.~\cite{Baak:2014wma}.
The two free parameters of the fit, namely the normalisations of the
\ttbar{} and \wjets{} background components, are constrained by the number of
events observed in the control regions and signal region sidebands, where the number of events is
described by a Poisson probability density function. The remaining
nuisance parameters correspond to the sources of systematic uncertainty
described in Sect.~\ref{sec:systematics}.
They are taken into account with their
uncertainties, and adjusted to maximise the likelihood.  The yields
estimated with the background\hyp only fit are reported in
Table~\ref{tab:results_blinded_bkgOnlyFit}, as well as the resulting predictions in \SRlbbone\ and \SRlbbtwo\ for $105<\mbb<135$ GeV.
While \CRlbbT{} is dominated by \ttbar{} events, \CRlbbW{} is populated evenly by \ttbar{} and \wjets{} events,
which causes the normalisations of the \ttbar{} and \wjets{}
contributions to be negatively correlated after the fit.
As a result, the uncertainties on individual background
sources do not add up quadratically to the uncertainty on the total SM
expectation. The normalisation factors are found to be $1.03\pm0.15$ for \ttbar{} and $0.79\pm0.07$ for \wjets{},
where the errors include statistical and systematic uncertainties.

To validate the background modelling, two validation regions (VR)
are defined similarly to the SRs except for requiring $40<\mtW<100\GeV$,
and requiring $100<\mct<160\GeV$ for \VRlbbone{} and $\mct>160\GeV$ for \VRlbbtwo{} as summarised in Table~\ref{tab:SRlbb}.
The yields in the VRs are shown in Table~\ref{tab:results_blinded_bkgOnlyFit} after the background\hyp only fit, which does not use the data in the VRs to constrain the background.
The data event yields are found to be consistent with background expectations.
Figure~\ref{fig:SRlbb} shows the data distributions of \mct, \mtW, \nbjet\ and \mbb\ compared to the SM expectations
in various regions.
The data agree well with the SM expectations in all distributions.

%%%%%%%%%%%%%%%%%%%%%%%%%%%%%%%%%%%%%%%%%%%%%%%

\section{One lepton and two photons channel}
\label{sec:lgg}

\subsection{Event Selection}

Events recorded with diphoton or single-lepton triggers are used in the one lepton and two photons channel.
For the diphoton trigger, the transverse momentum thresholds at trigger level for the highest-\pt\ (leading) 
and second highest-\pt\ (sub-leading) photons are $35\GeV$ and $25\GeV$ respectively.
For these events, the event selection requires exactly one signal lepton ($e$ or $\mu$)
and exactly two signal photons, with \pt\ thresholds of $15\GeV$ for electrons,
$10\GeV$ for muons, and 40 (27) GeV for leading (sub-leading) photons.  
In addition, events recorded with single-lepton triggers, which have  transverse momentum thresholds at trigger level of 24 GeV, are used.  For these events, the selection requires \pt\ thresholds of $25\GeV$ for electrons and muons, and 40 (20) GeV for leading (sub-leading) photons. 

\begin{table}[b]
\centering
\caption{\label{tab:SRlgg}
Selection requirements for the signal and validation regions of the one lepton and two photons channel.
The number of leptons and photons is labelled with \nlep\ and \ngamma\ respectively.}
\begin{tabular}{lcccc}
\toprule
              & \SRlggone & \SRlggtwo & \VRlggone & \VRlggtwo \\
\midrule
\nlep         & 1 & 1 & 1 & 1 \\
\ngamma       & 2 & 2 & 2 & 2 \\
\met~[GeV]    & $>40$  & $>40$   & $<40$ & --- \\
\DphiWh & $>2.25$& $>2.25$ & ---& $<2.25$ \\
\mtgone [GeV] & $>150$ & $<150$ & & \\
              & and    & or & --- & ---\\
\mtgtwo [GeV] & $> 80$ & $< 80$ & & \\
\bottomrule
\end{tabular}
\end{table} 
\begin{figure*}[t!]\centering
\subfigure[\met\ in \SRlggone\ and \SRlggtwo\ without \met\ cut]
	{\includegraphics[width=0.48\textwidth]{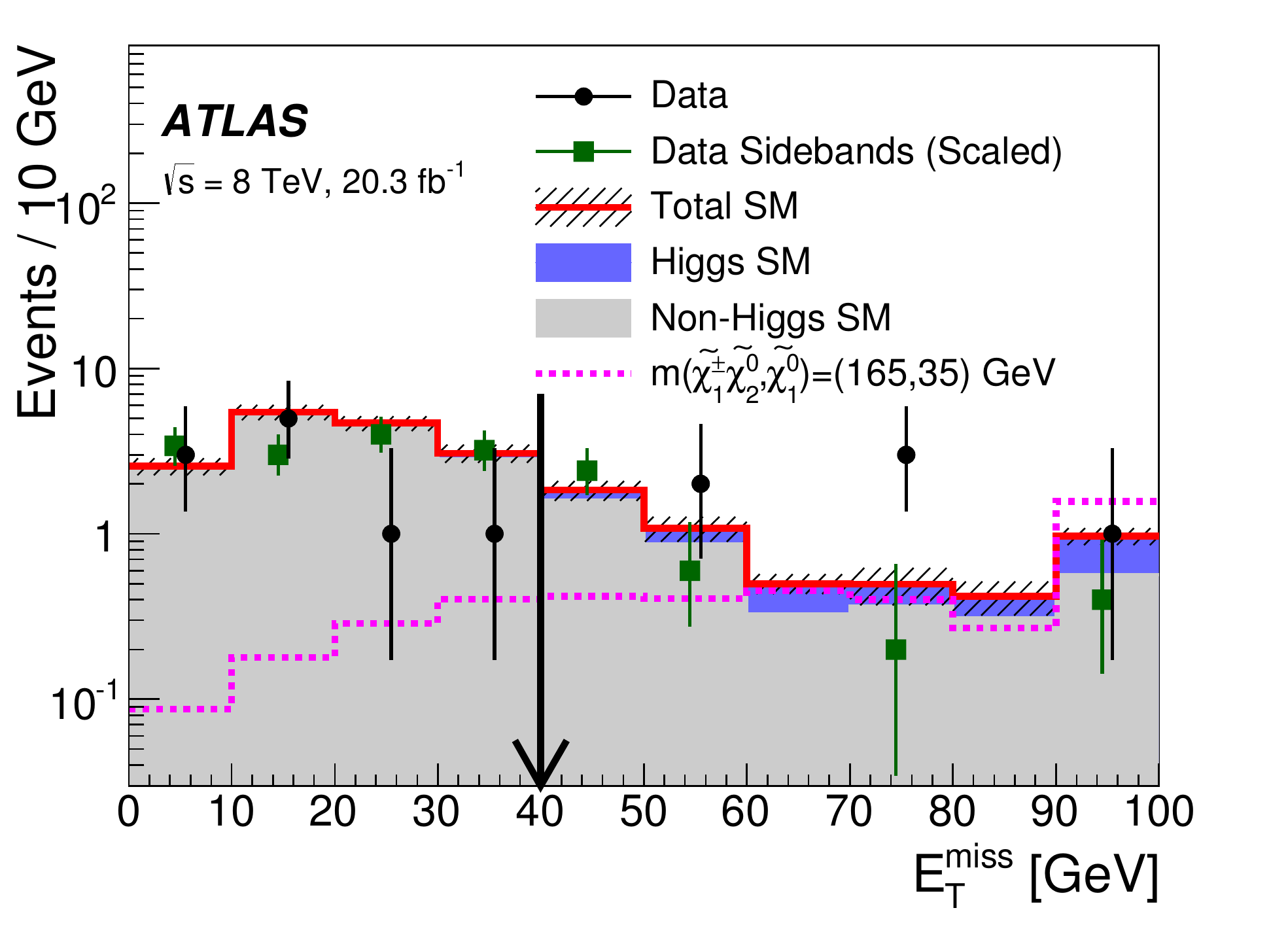}}\hfil
\subfigure[\DphiWh\ in \SRlggone\ and \SRlggtwo\ without \DphiWh\  cut]
	{\includegraphics[width=0.48\textwidth]{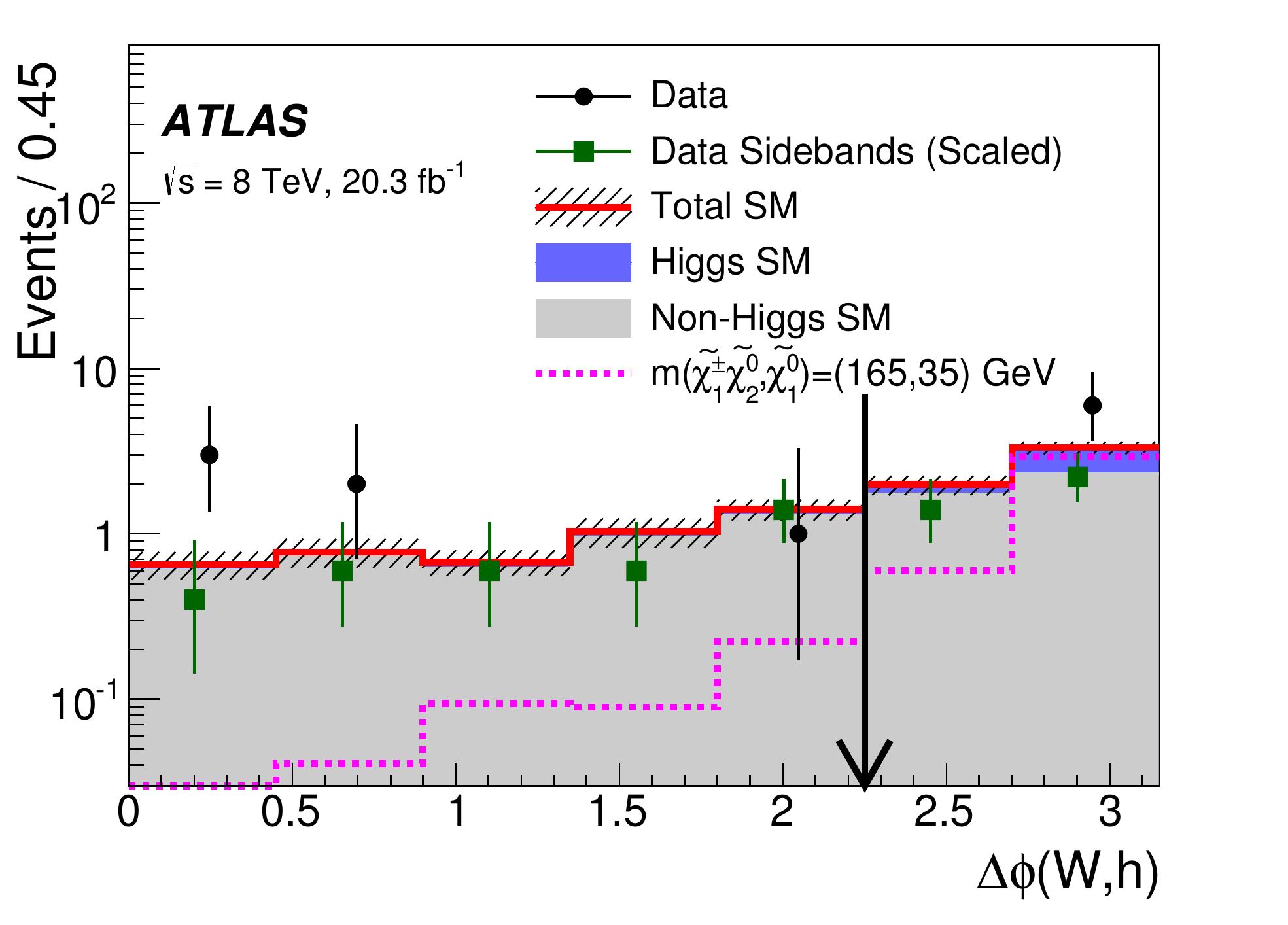}}
\subfigure[\mtgone\ in \SRlggone\ and \SRlggtwo\ without $m_{\mathrm{T}}^{\smash{W\!\gamma_i}}$ cuts]
	{\includegraphics[width=0.48\textwidth]{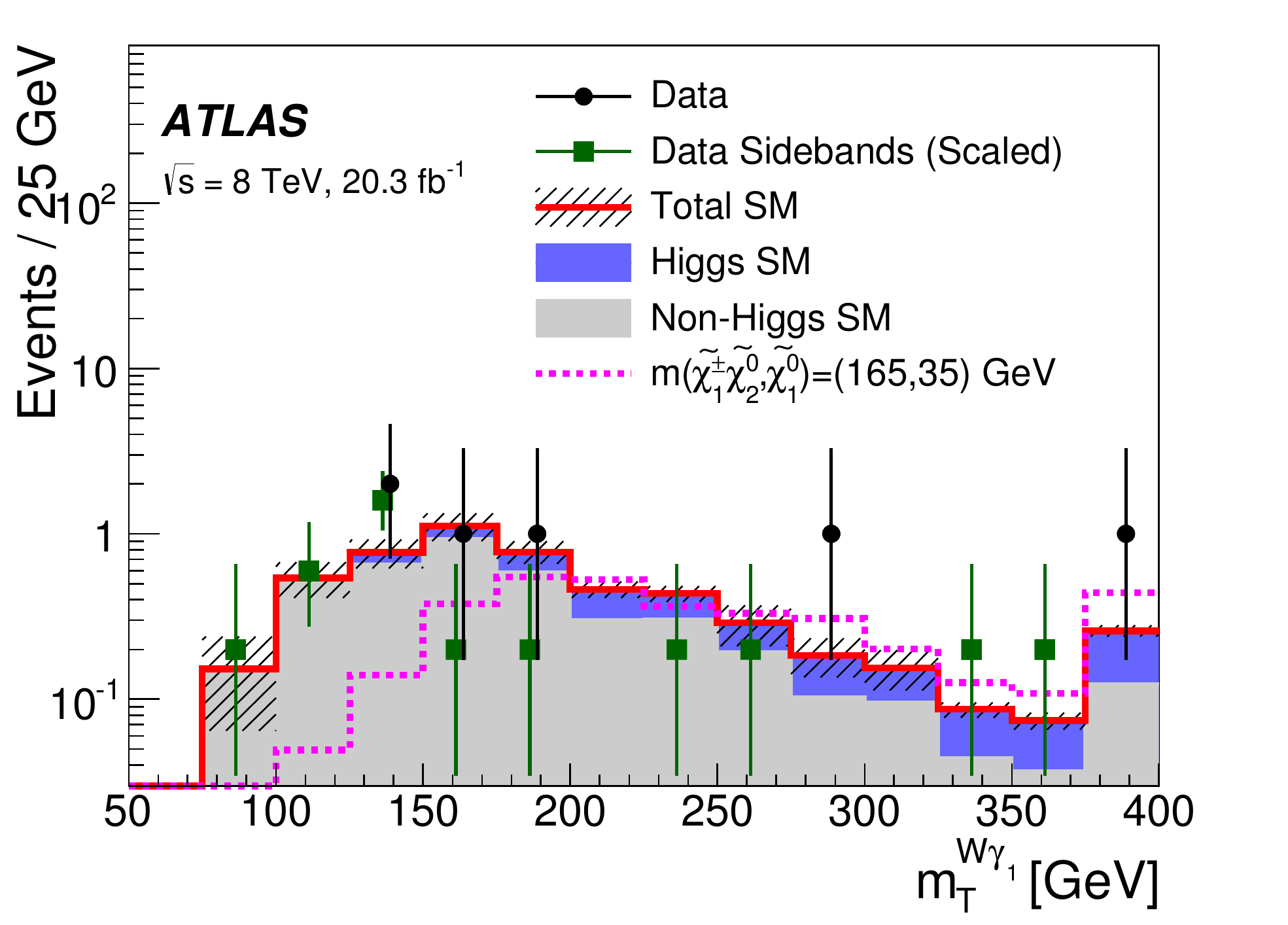}}\hfil
\subfigure[\mtgtwo\ in \SRlggone\ and \SRlggtwo\ without $m_{\mathrm{T}}^{\smash{W\!\gamma_i}}$ cuts]
	{\includegraphics[width=0.48\textwidth]{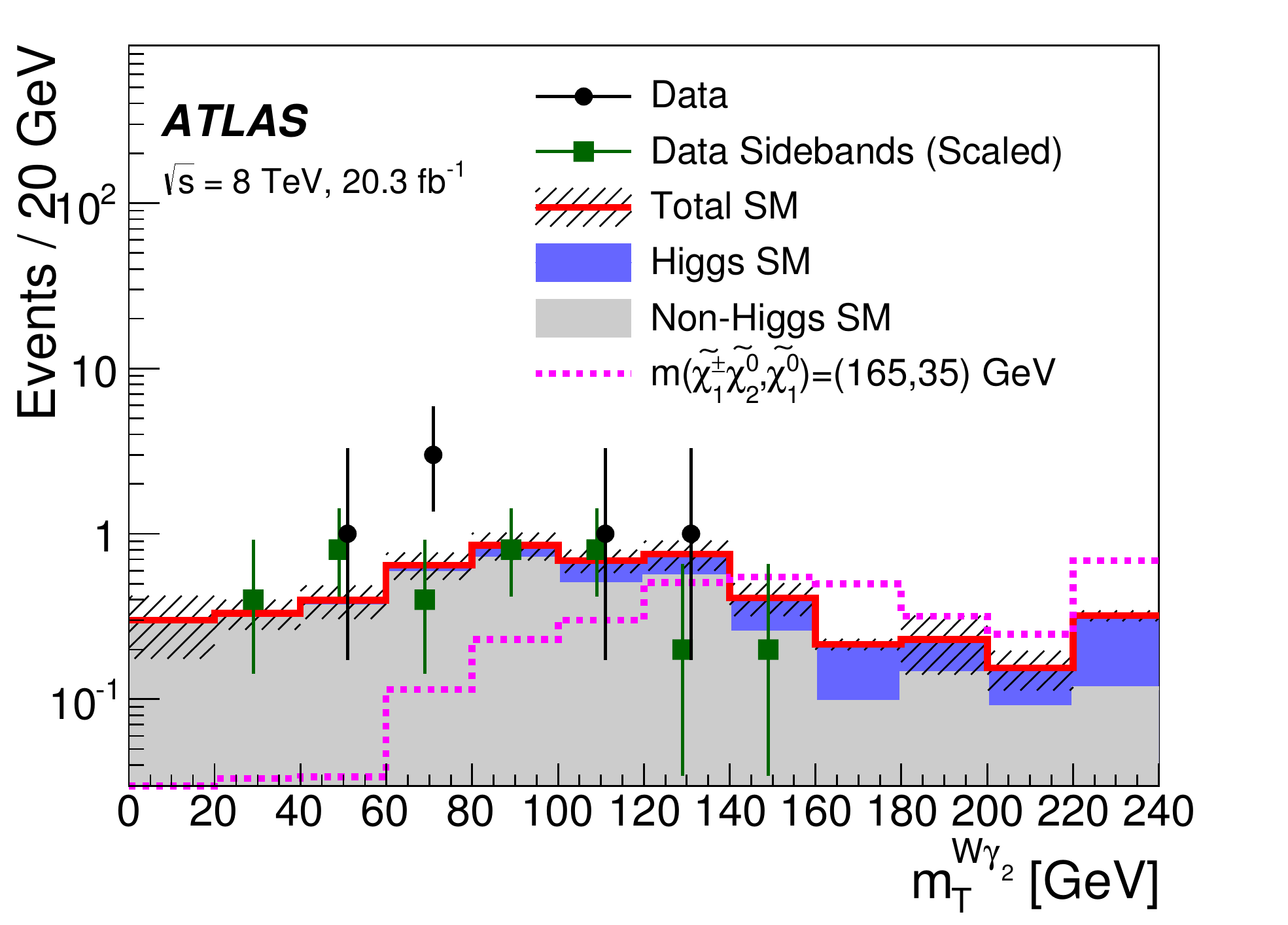}}
\caption{\label{fig:SRlggvars}
	Distributions of missing transverse momentum \met, azimuth difference between the $W$ and Higgs boson candidates \DphiWh, transverse mass of the $W$ and photon system \mtgone\ and \mtgtwo{} in the one lepton and two photons signal regions for the Higgs-mass window ($120<\mgg<130\GeV$). 
The vertical arrows indicate the boundaries of the signal regions.
The filled and hashed areas represent the stacked histograms of the simulation-based background cross check and the total uncertainties.  The contributions from non-Higgs backgrounds are scaled by 10 \GeV\ / 50 \GeV\ = 0.2 from the \mgg{} sideband ($100<\mgg<120\GeV$ and $130<\mgg<160\GeV$) into the Higgs-mass window.
	The rightmost bins in (a), (c), and (d) include overflow.  Scaled data in the sideband  are shown as squares, while events in the Higgs-mass window are shown as circles.  
The distributions of a signal hypothesis are also shown without stacking on the background histograms.}
\end{figure*}

\begin{figure*}[ht!]\centering
\subfigure[\SRlggone]
	{\includegraphics[width=0.48\textwidth]{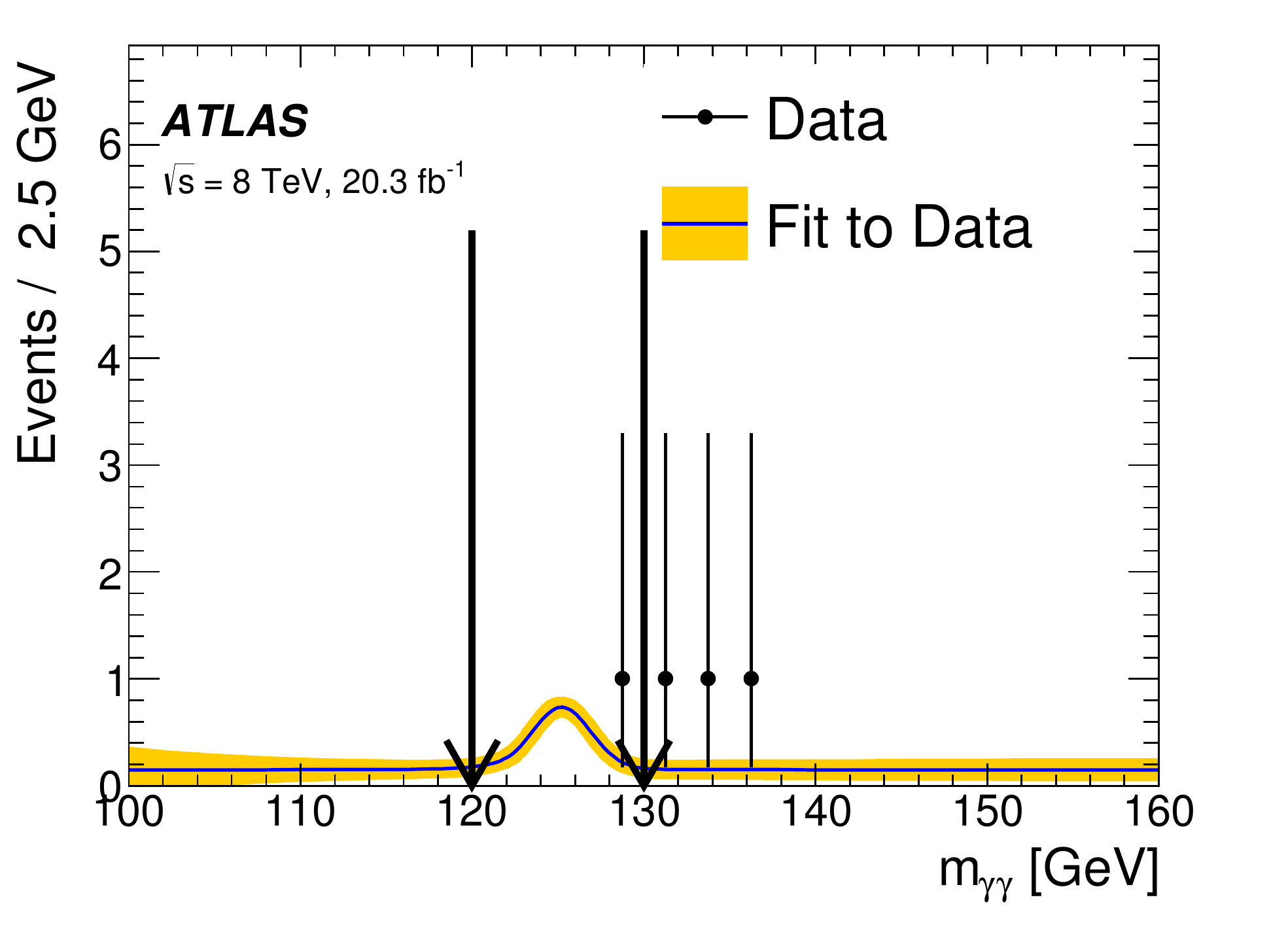}}\hfil
\subfigure[\SRlggtwo]
	{\includegraphics[width=0.48\textwidth]{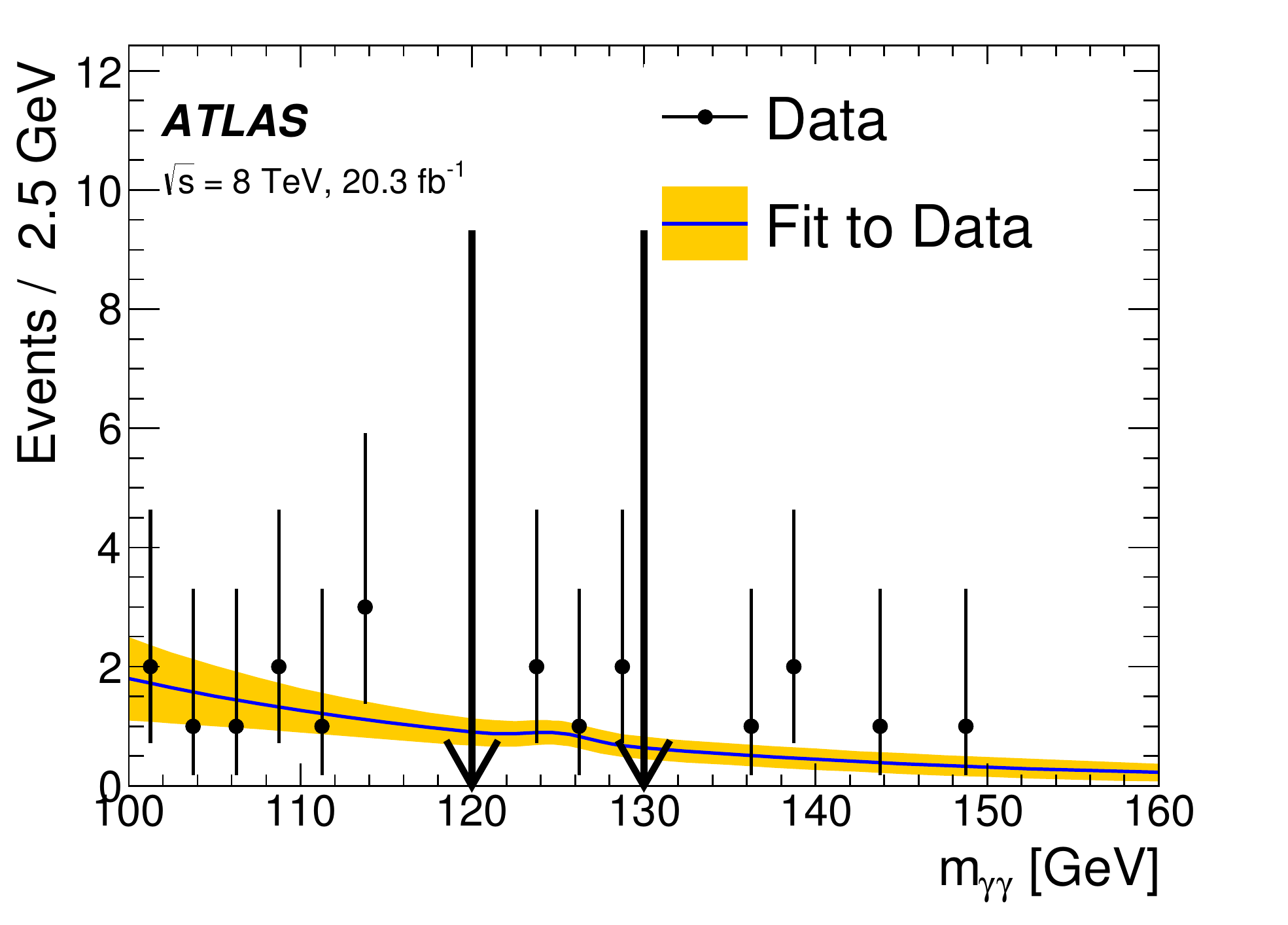}}
\subfigure[\VRlggone]
	{\includegraphics[width=0.48\textwidth]{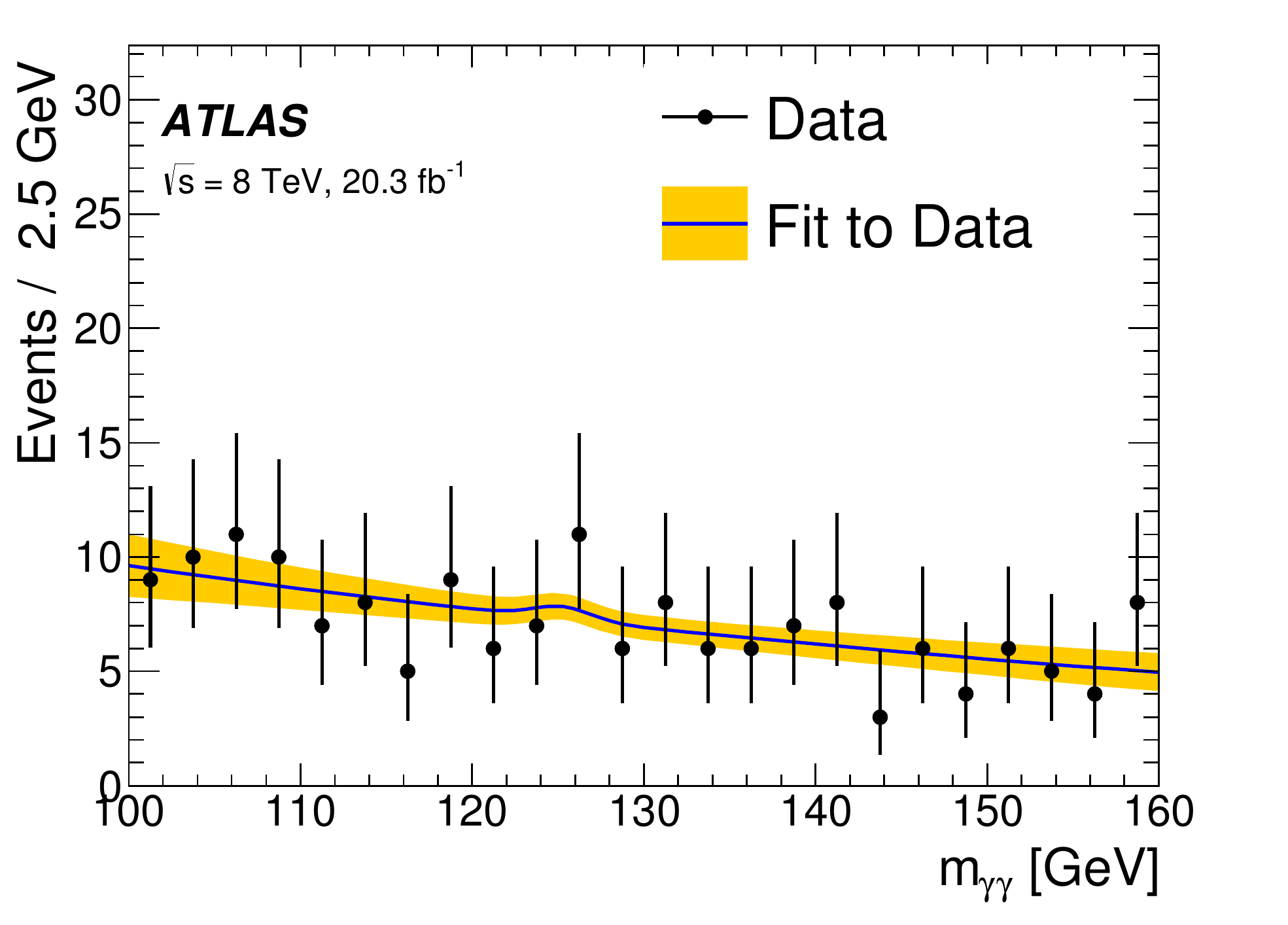}}\hfil
\subfigure[\VRlggtwo]
	{\includegraphics[width=0.48\textwidth]{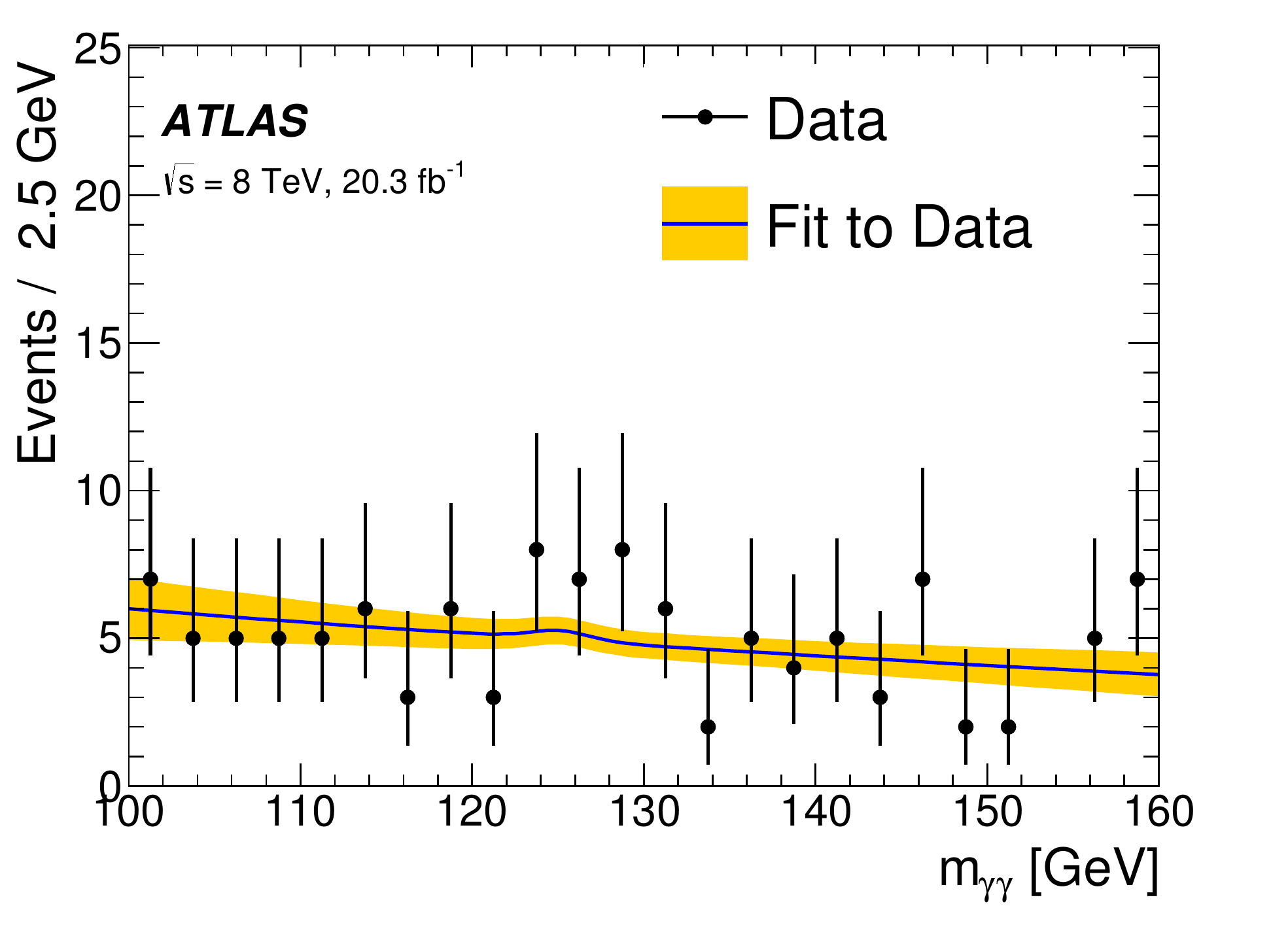}}
\caption{\label{fig:SRlgg}
	Results of the background\hyp only fit to the diphoton invariant mass, \mgg, distribution in the one lepton and two photons signal
	and validation regions.  The contributions from SM Higgs boson production are constrained to the MC prediction and associated systematic uncertainties.
	The band shows the systematic uncertainty on the fit.  The fit is performed on events with 100 GeV $< \mgg <$ 160 GeV, with events in 
	\SRlggone{} or \SRlggtwo{} in the Higgs-mass window (120 GeV $\le \mgg \le$ 130 GeV), indicated by the arrows, excluded from the fit.
}
\end{figure*}

In this channel, 
a neural network algorithm, based on the
momenta of the tracks associated with each vertex and the direction
of flight of the photons, is used to select the primary
vertex, similarly to the ATLAS SM $h\to\gamma\gamma$
analysis described in Ref.~\cite{ATLAS:higgs2013}.  
Signal muons must satisfy $|d_0|<1$\,mm and $|z_0|<10$\,mm.
The isolation criteria for both the electrons and muons are
$\etcone{0.4}/\pt<0.2$ and $\ptcone{0.2}/\pt<0.15$.
Signal photons are required to satisfy $\etcone{0.4}<6\GeV$ and $\ptcone{0.2}<2.6\GeV$.

The two largest background contributions are due to multi-jet
and $Z\gamma$ production, with leptons
or jets misreconstructed as photons.
These background contributions are suppressed by requiring  $\met>40\GeV$.

\begin{sloppypar}
The $\pTvec$ of the $W\to\ell\nu$ system, reconstructed assuming background events with neutrino $\pTvec=\pTmiss$, is required to be back-to-back with the $\pTvec$ of the $h\to\gamma\gamma$ candidate ($\DphiWh>2.25$).
Only events with a diphoton invariant mass, \mgg, between 100 and 160~GeV are 
considered. Events in the sideband, outside the Higgs-mass window between 120 and 130~GeV, are included to constrain the non-Higgs background as described in Sect.~\ref{sec:lyy-bkg}.
\end{sloppypar}

Selected events are split into two SRs with different expected signal sensitivities
based on two variables \mtgone\ and \mtgtwo, which are defined as
\begin{equation}
m_\mathrm{T}^{\smash{W\!\gamma_i}} = \sqrt{ (\mtW)^2 + 2 \et^W \et^{\gamma_i} - 2 \pTvec^W\cdot\pTvec^{\gamma_i}},
\end{equation}
where $\mtW$, $\et^W$ and $\pTvec^W$ are the transverse mass, energy and momentum of the $W$ candidate,
and $\et^{\gamma_i}$ and $\pTvec^{\gamma_i}$ are the transverse energy and momentum of the $i$-th, $\pt$-ordered, photon.
Including a photon in the transverse mass calculation provides a means to identify
leptonically decaying $W$ bosons in the presence of a final-state radiation photon.
Events with $\mtgone>150\GeV$ and $\mtgtwo>80\GeV$ are classified into
\SRlggone, and those with either $\mtgone<150\GeV$ or $\mtgtwo<80\GeV$
into \SRlggtwo. Most of the sensitivity to the signal is provided by \SRlggone{}, 
while \SRlggtwo{} assists in constraining systematic uncertainties.

Two overlapping validation regions are defined by 
inverting and modifying the \met\ and $\DphiWh$ criteria relative to
those of the signal regions.
The first region \VRlggone\ requires $\met<40\GeV$ and has no requirement on \DphiWh,
and the second region \VRlggtwo\ requires $\DphiWh<2.25$ and has no requirement on \met.
The signal and validation regions are summarised in Table~\ref{tab:SRlgg}.

\begin{table*}[ht!]
\caption{\label{tab:lggbackground}
Event yields and SM expectation in the Higgs-mass window of the lepton plus two photon channel ($120<\mgg<130\GeV$) after the background\hyp only fit.
The Higgs-mass window is excluded from the fit in the two signal regions.
The errors shown include statistical and systematic uncertainties.}
\begin{center}
\begin{tabular}{ll@{\,$\pm$\,}ll@{\,$\pm$\,}ll@{\,$\pm$\,}ll@{\,$\pm$\,}l}
\toprule
& \multicolumn{2}{c}{\SRlggone}
& \multicolumn{2}{c}{\SRlggtwo}
& \multicolumn{2}{c}{\VRlggone}
& \multicolumn{2}{c}{\VRlggtwo}\\
\midrule
Observed events & \multicolumn{1}{l}{1}& & \multicolumn{1}{l}{5}& & \multicolumn{1}{l}{30}& & \multicolumn{1}{l}{26}& \\
SM expectation  & 1.6 &0.4  & 3.3 &0.8 & 30.2 &2.3  & 20.4 &1.9  \\
\midrule
Non-Higgs       & 0.6 &0.3  & 3.0 &0.8 & 29.2 &2.3  & 19.8 &1.9  \\
$Wh$            & 0.85&0.02 & 0.23&0.01& 0.71 &0.02 & 0.29 &0.01 \\
$Zh$            & 0.04&0.01 & 0.02&0.01& 0.14 &0.02 & 0.05 &0.01 \\
\tth            & 0.14&0.01 & 0.02&0.01& 0.11 &0.01 & 0.25 &0.01 \\
\bottomrule
\end{tabular}
\end{center}
\end{table*}

\begin{sloppypar}
Distributions in the Higgs-mass window of the four kinematic variables used to define the
SRs are shown in Fig.~\ref{fig:SRlggvars}.  For illustration purposes, the observed yield in the sideband region is shown for each distribution, scaled into the corresponding Higgs-mass
window by the relative widths of the Higgs-mass window and the sideband region, 10 \GeV\ / 50 \GeV\ = 0.2.
Also shown, for each distribution, is a simulation-based cross-check of the background estimate.
To reduce statistical uncertainties originating from the limited number of simulated events, the non-Higgs contributions are obtained in the sideband and scaled into the Higgs-mass window by 0.2.
The simulation-based prediction of the non-Higgs background is estimated from the
W/Z($\gamma,\gamma\gamma$) +jets samples, after applying a data-driven
correction for the probability of
electrons or jets to be reconstructed as photons.  The contribution from
backgrounds with jets reconstructed as leptons is determined by using the ``fake factor''
method described in Ref.~\cite{PhysRevD.87.052002}. 
This simulation-based background estimate is only used as a
cross-check of the sideband-data-based background estimate described above.
It gives results consistent with the data estimate, but it is not used
for limit setting.
\end{sloppypar}

%%%%%%%%%%%%%%%%%%%%%%%%%%%%%%%%%%%%%%%%%%%%%%%

\subsection{Background estimation}
\label{sec:lyy-bkg}

The contribution from background sources
that do not contain a $h\to\gamma\gamma$ decay can be statistically separated 
by a template fit to the full \mgg\ distribution, from 100~GeV to 160~GeV.
The approach followed is similar to the one in Ref.~\cite{ATLAS:higgs2013}:
the non-Higgs background is modelled as $\exp(-\alpha\mgg)$, with the
constant $\alpha$ as a free, positive parameter in the fit.
Alternative functional models are used to evaluate the systematic uncertainty due to the choice of background modelling function.
The $h\to\gamma\gamma$ template, used for the Higgs background and signal, 
is formed by the sum of a Crystal Ball function~\cite{Oreglia:1980cs} for the core of the distribution and a Gaussian function for the tails.
 This functional form follows the one used in the SM $h\to\gamma\gamma$ analysis~\cite{ATLAS:higgs2013},
with the nominal values and uncertainties on the fit parameters determined by
fits to the simulation in \SRlggone{} and \SRlggtwo{}. The results of the fit to the simulation are used as an external constraint 
on the template during the fit to data.
The width of the Gaussian core of the Crystal Ball function quantifies the 
detector resolution and is determined in simulation to be 1.7 \GeV\ in \SRlggone\ and 1.8 \GeV\ in \SRlggtwo.  This is comparable to the resolution found 
in the SM $h\to\gamma\gamma$ analysis~\cite{ATLAS:higgs2013}.

Contributions from SM processes with a real Higgs boson decay are estimated by simulation and come primarily from 
$Wh$ associated production, with smaller amounts from $\ttbar h$ and $Zh$.  The contributions from SM Higgs boson production 
via gluon fusion or vector boson fusion are found to be negligible.  Systematic uncertainties on the yields of 
these SM processes are discussed in Sect.~\ref{sec:systematics}.
Figure~\ref{fig:SRlgg} shows the background\hyp only fits to the observed \mgg\ distributions in
the signal and validation regions, with the signal region Higgs-mass window 
($120<\mgg<130\GeV$) excluded from the fit.
Table~\ref{tab:lggbackground} summarises the observed event yields
 in the Higgs-mass window and the background estimates, from the 
background\hyp only fits, in the signal and validation regions.
The errors are dominated by the statistical uncertainty due to the
 number of events in the \mgg\ sidebands.

%%%%%%%%%%%%%%%%%%%%%%%%%%%%%%%%%%%%%%%%%%%%%%%

\section{Same-sign dilepton channel}
\label{sec:ss-2l}

\subsection{Event Selection}
\label{sec:ss-2l-sel}

\begin{sloppypar}
Events recorded with a combination of dilepton triggers are used in the same-sign dilepton channel.
The \pt{} thresholds of the dilepton triggers depend on the flavour of the leptons.
The triggers reach their maximum efficiency at \pt{} values of
about $14-25$~\GeV{} for the leading lepton and $8-14$~\GeV{} for
the sub-leading lepton.

\begin{table*}[ht!]\centering
\caption{\label{tab:SRll}
Selection requirements for the signal regions of the same-sign dilepton channel.}
\begin{tabular}{ccccccc}
\toprule
& \SReeone & \SReetwo & \SRmmone & \SRmmtwo & \SRemone & \SRemtwo \\
\midrule
Lepton flavours              & $ee$   & $ee$   & $\mu\mu$ & $\mu\mu$ & $e\mu$ & $e\mu$ \\
\njet               & 1      & 2 or 3 & 1        & 2 or 3   & 1      & 2 or 3 \\
Leading lepton \pt\ [GeV]    & $>30$  & $>30$  & $>30$    & $>30$    & $>30$  & $>30$  \\
Sub-leading lepton \pt\ [GeV] & $>20$  & $>20$  & $>20$    & $>30$    & $>30$  & $>30$  \\
$|\mll-m_Z|$ [GeV]           & $>10$  & $>10$  & --      & --      & --    & --    \\
\detall\                     & --    & --    & $<1.5$   & $<1.5$   & $<1.5$ & $<1.5$ \\
\metrel\ [GeV]               & $>55$  & $>30$  & --      & --      & --    & --    \\
\meff{}  [GeV]               & $>200$ & --    & $>200$   & $>200$   & $>200$ & $>200$ \\
\mtmax{} [GeV]               & --    & $>110$ & $>110$   & --      & $>110$ & $>110$ \\
\mlj\ or \mljj\ [GeV]        & $<90$  & $<120$ & $<90$    & $<120$   & $<90$  & $<120$ \\
\bottomrule
\end{tabular}
\end{table*}

The offline event selection requires two same-sign signal leptons ($ee$, $e\mu$ or $\mu\mu$) with
$\pt>30\GeV$ or $20\GeV$ as shown in Table~\ref{tab:SRll} and no additional preselected lepton.
The signal electrons must satisfy the ``tight'' identification criteria from Ref.~\cite{Aad:2014fxa},
$|d_0|/\sigma_{d_0}<3$, and $|z_0\sin\theta{}|<0.4$\,mm.
The signal muons must satisfy $|\eta|<2.4$, $|d_0|/\sigma_{d_0}<3$, and $|z_0\sin\theta{}|<1$\,mm.
The isolation criteria for electrons (muons) are 
$\etcone{0.3}/$$\min(\pt,60\GeV)$$<0.13$ (0.14) and $\ptcone{0.3}/$$\min(\pt,60\GeV)$$<0.07$ (0.06).
Events containing a hadronically decaying preselected $\tau$ lepton are
rejected in order to avoid statistical overlap with the three-lepton
final states~\cite{ATLAS:3L8TeV}.
\end{sloppypar}

Events are required to contain one, two, or three central
($|\eta|<2.4$) jets with $\pt>20\GeV$. If a central jet
has $\pt{}<50\GeV$ and has tracks associated to it, at least one of
the tracks must originate from the event primary vertex.
To reduce background contributions with heavy-flavour decays, all the jets must fail to meet the $b$-tagging criterion at the 80\% efficiency operating point.
There must be no forward ($2.4<|\eta|<4.9$) jet with $\pt>30\GeV$.

\begin{sloppypar}
The dominant background contributions in the \llss{} channel are due to SM diboson
production ($WZ$ and $ZZ$) leading to two ``prompt" leptons and due to events with ``non-prompt" leptons
(heavy-flavour decays, photon conversions and misidentified jets). These
background contributions are suppressed with the tight identification criteria
described above, and with the kinematic requirements summarised in
Table~\ref{tab:SRll}. The requirements were optimised separately
for each lepton flavour combination ($ee$, $\mu\mu$, and $e\mu$), and
for different numbers of reconstructed jets, leading to six signal regions.
\end{sloppypar}

The dilepton invariant mass \mll\ is required to differ by at least 10\,GeV from the $Z$-boson mass
for the $ee$ channel, in which contamination due to electron charge misidentification is significant.

The visible mass of the Higgs boson candidate is defined for the one jet signal regions as the invariant mass (\mlj) of the jet and the lepton that is closest to it in terms of $\Delta R$, and for the two or three jet signal regions as the invariant mass (\mljj) of the two highest-\pt\ jets and the lepton that is closest to the dijet system.
In the signal regions, $\mlj{}<90$~\GeV{} is required for \SRllone{}
and $\mljj{}<120$~\GeV{} for \SRlltwo{}.

\begin{figure}[b!]
\centering
\includegraphics[width=0.45\textwidth]{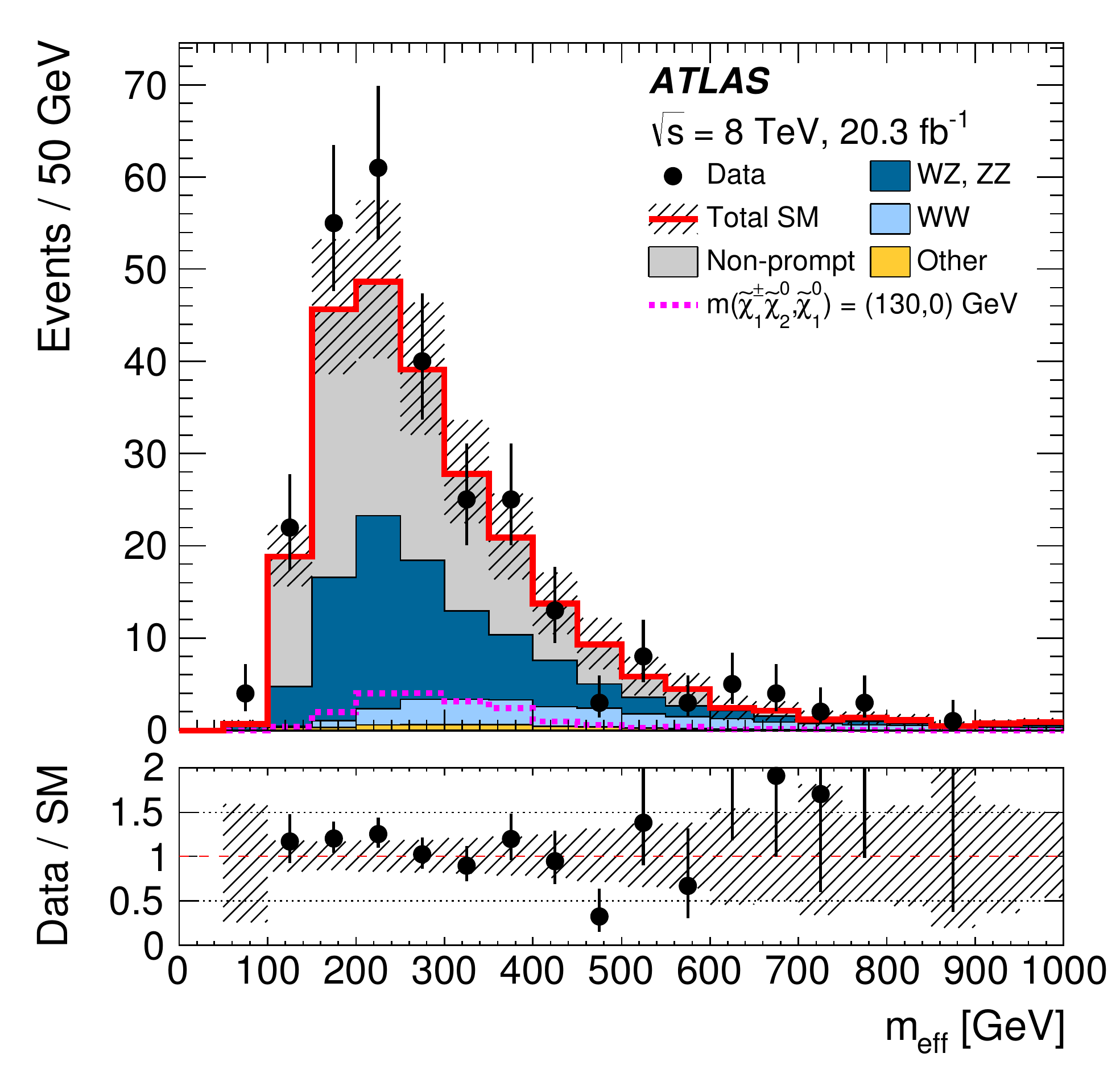}
\caption{\label{fig:fakeVRll} Distribution of effective mass \meff{} in the validation
  region of the same-sign $e\mu$ channel. This validation region is
  defined by requiring one, two, or three jets, and reversing the \mlj{},
  \mljj{} criteria. The hashed areas represent the total
  uncertainties on the background estimates that are depicted with stacked histograms. The distribution of a
  signal hypothesis is also shown without stacking on the background histograms. The lower panel shows the ratio
  of the data to the SM background prediction.
}
\end{figure}

\begin{figure*}[p]
\centering
\subfigure[\meff{} in \SRllone\ without \meff\ cut]
	{\includegraphics[width=0.38\textwidth]{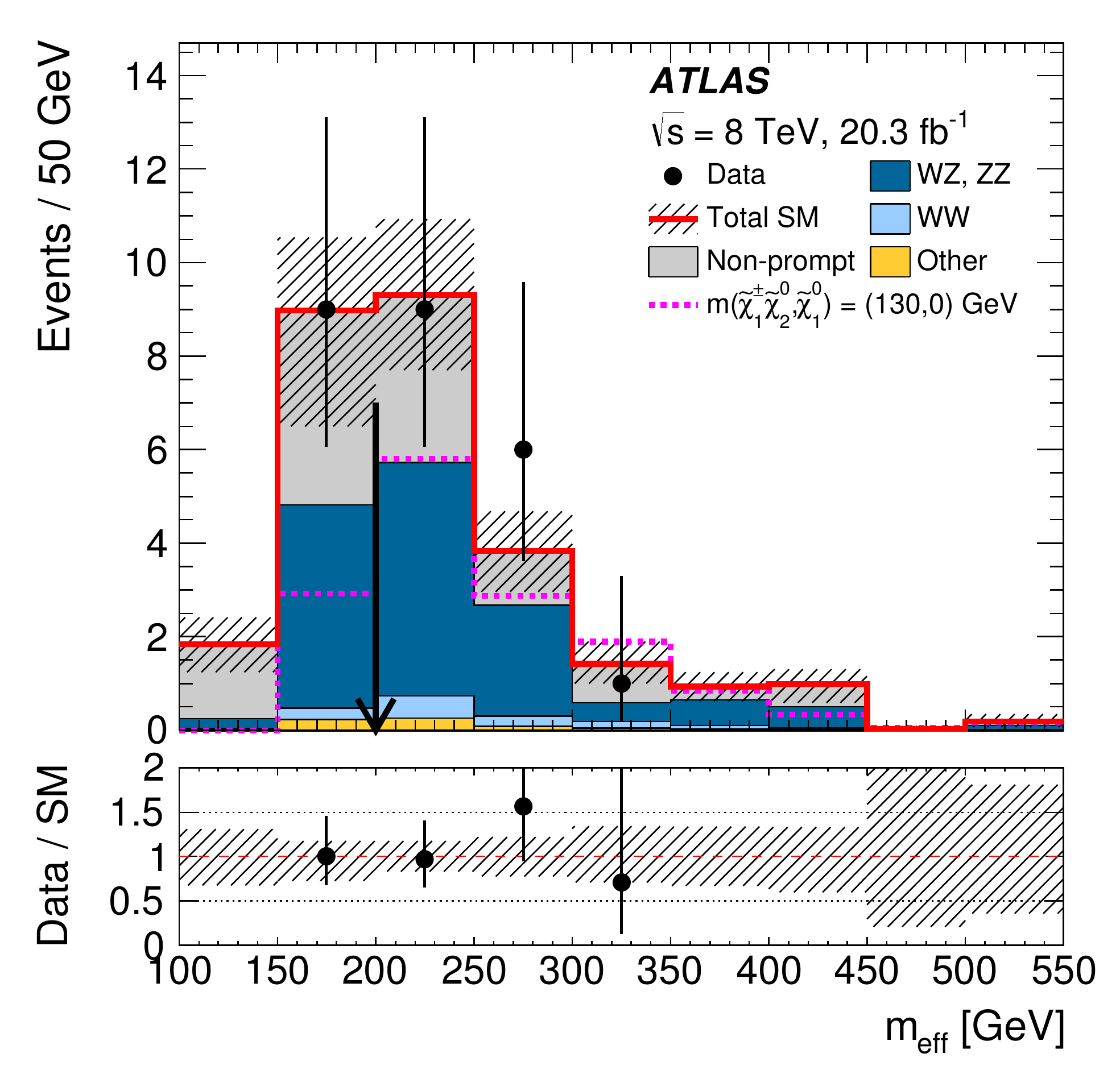}}\hspace{1.8cm}
\subfigure[\meff{} in \SRlltwo\ without \meff\ cut]
	{\includegraphics[width=0.38\textwidth]{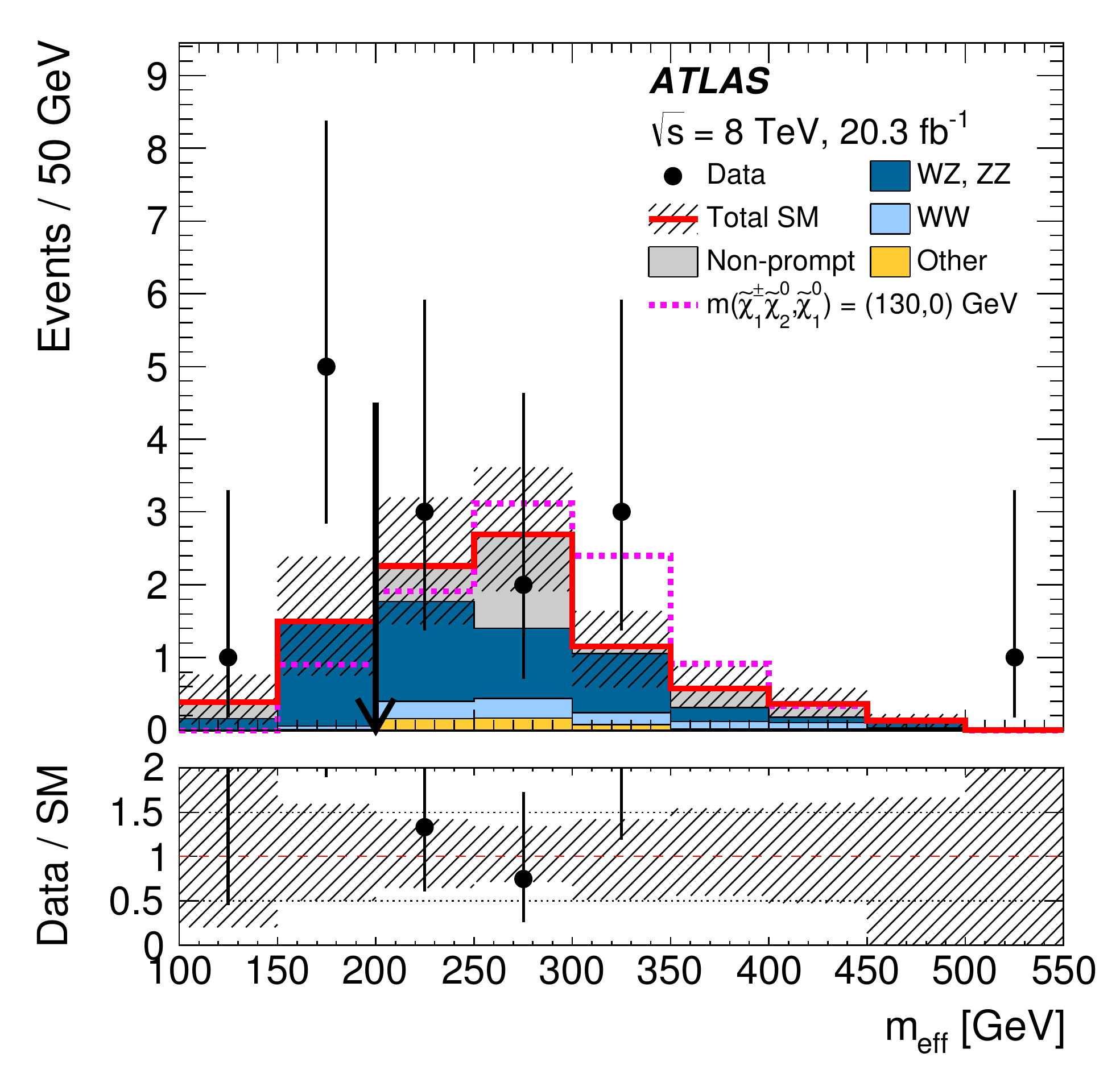}}
\subfigure[\mtmax\ in \SRllone\ without \mtmax\ cut]
	{\includegraphics[width=0.38\textwidth]{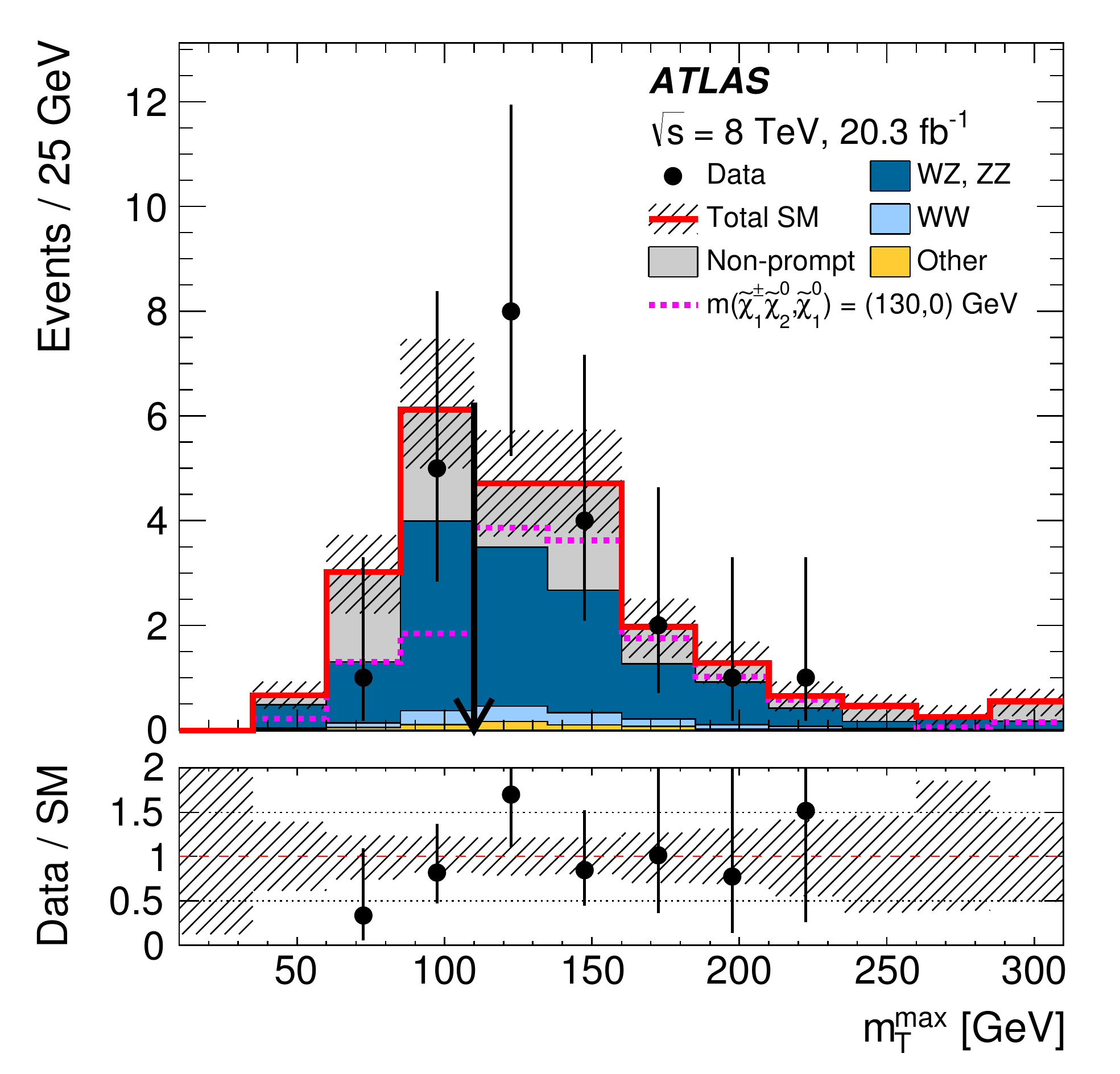}}\hspace{1.8cm}
\subfigure[\mtmax\ in \SRlltwo\ without \mtmax\ cut]
	{\includegraphics[width=0.38\textwidth]{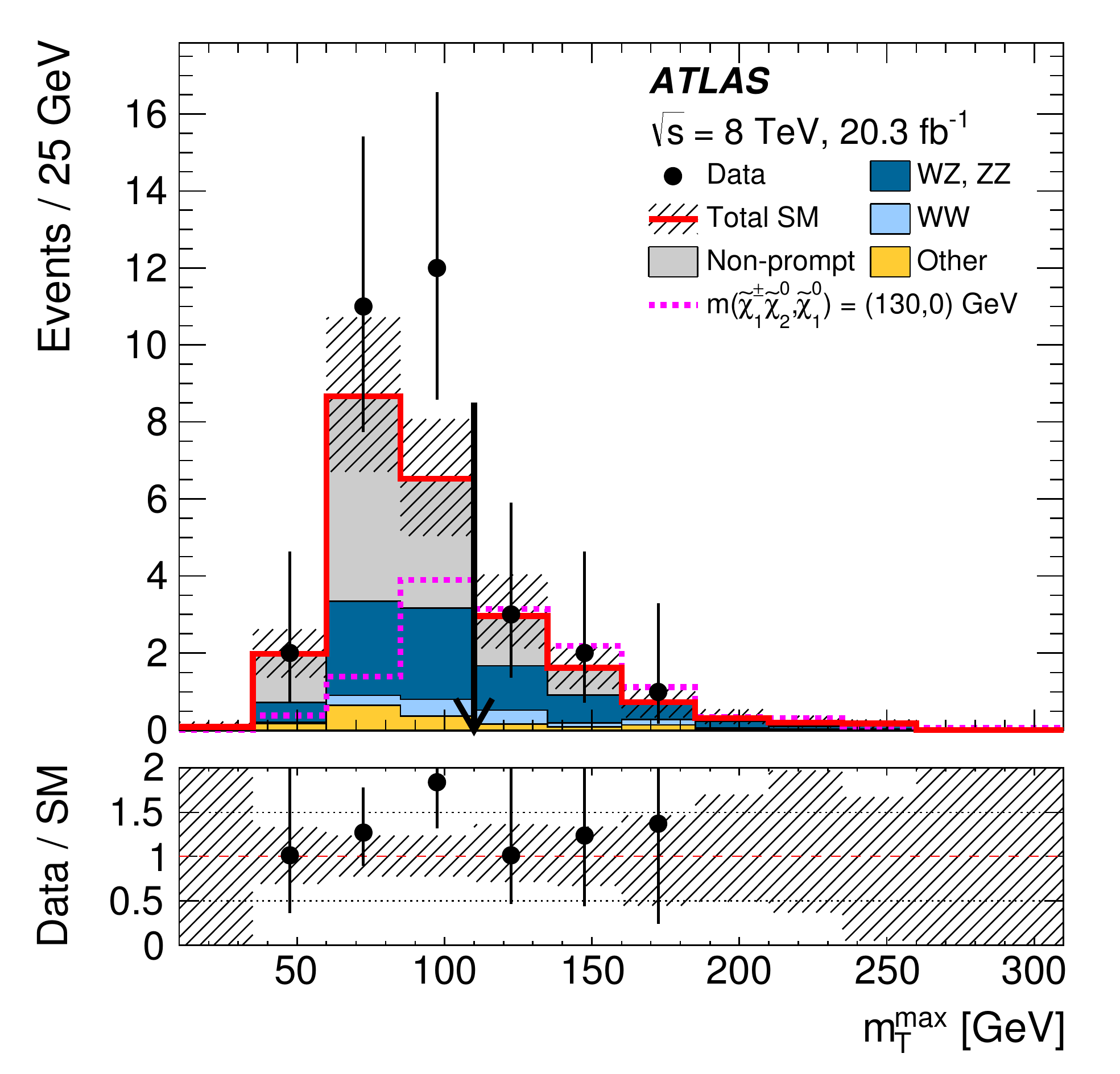}}
\subfigure[\mlj\ in \SRllone\ without \mlj\ cut]
	{\includegraphics[width=0.38\textwidth]{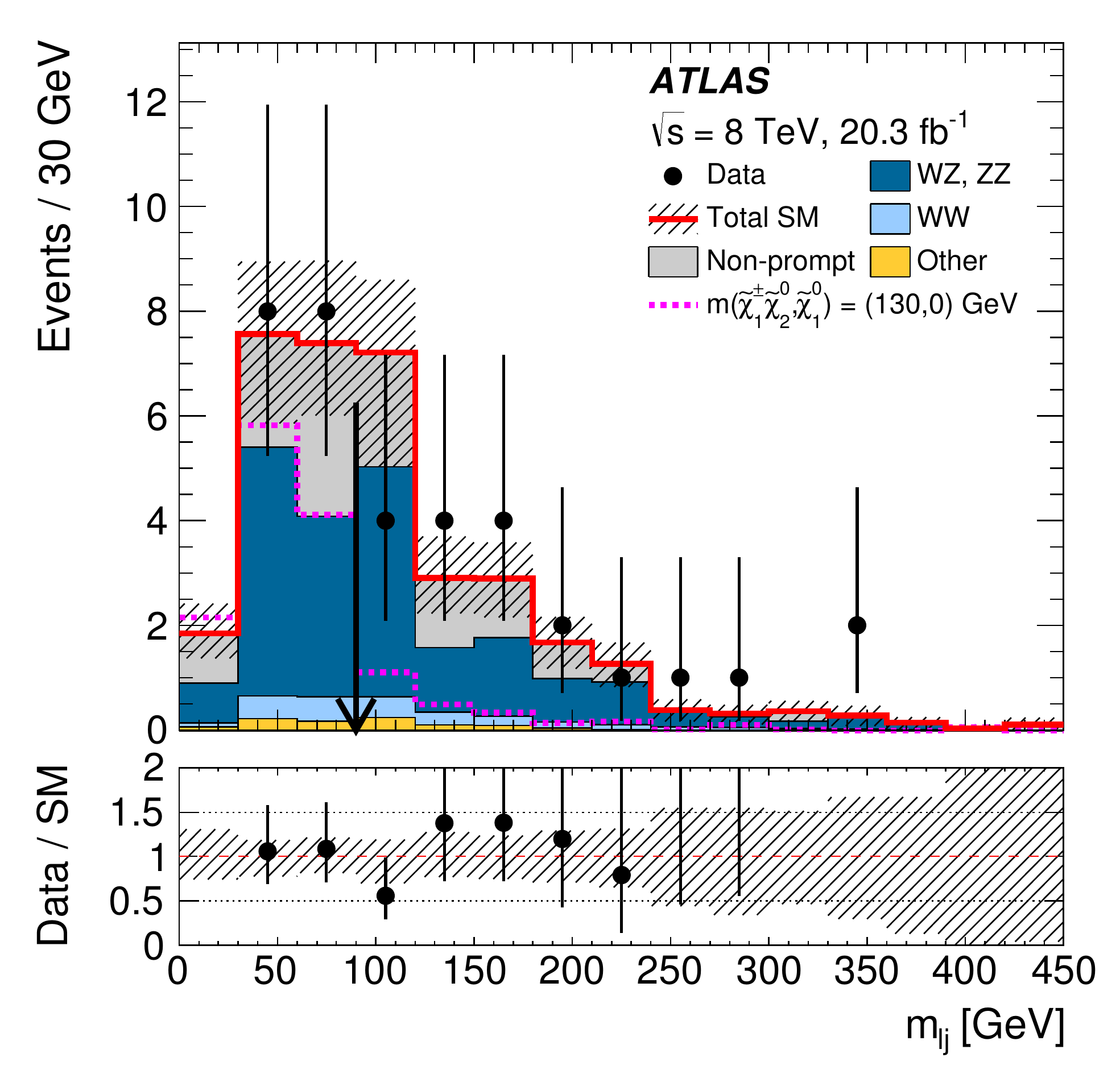}}\hspace{1.8cm}
\subfigure[\mljj\ in \SRlltwo\ without \mljj\ cut]
	{\includegraphics[width=0.38\textwidth]{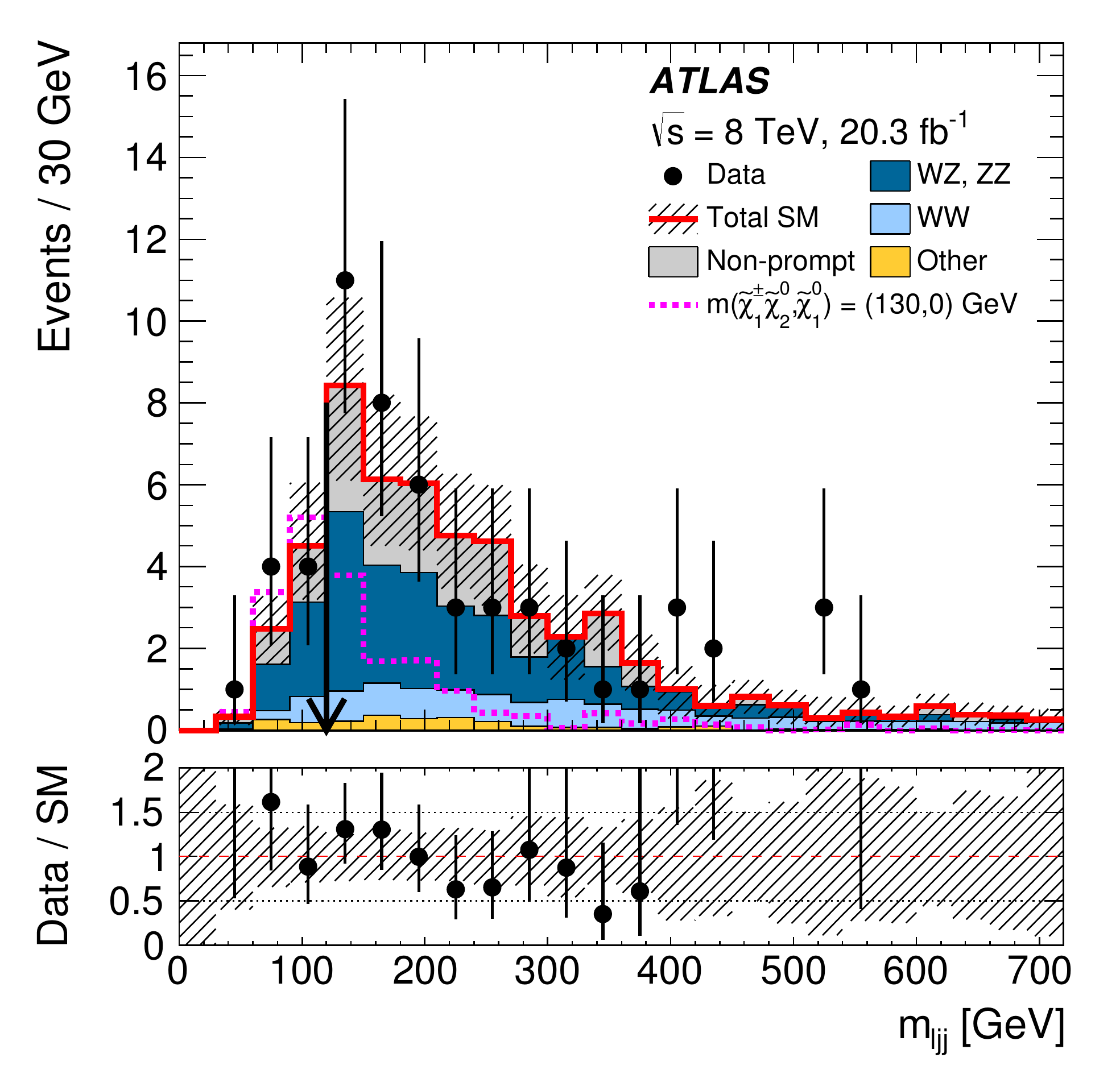}}
\caption{\label{fig:SRll}Distributions of effective mass \meff{}, largest transverse mass \mtmax{}, invariant mass of lepton and jets \mlj{} and
  \mljj{} for the same-sign dilepton channel in the signal regions with
  one jet (left) and two or three jets (right). \SRllone{} is the sum of
  \SReeone{}, \SRemone{}, and \SRmmone{}; \SRlltwo{} is the sum of
  \SReetwo{}, \SRemtwo{}, and \SRmmtwo{}. All selection criteria are
  applied, except for the one on the variable being shown.
  The vertical arrows indicate the boundaries of the signal regions, which may not apply to all flavour channels.
  The hashed areas represent the total
  uncertainties on the background estimates that are depicted with stacked histograms. The distributions of a
  signal hypothesis are also shown without stacking on the background histograms. The lower panels show the ratio
  between data and the SM background prediction. The rightmost bins of each distribution
  include overflow. 
  }
\end{figure*}

\begin{table*}[ht!]
\caption{
  \label{tab:ss2l_results_blinded_bkgOnlyFit}
  Event yields and SM expectation in the same-sign dilepton channel signal regions.
  The $WW$ background includes both $W^{\pm}W^{\pm}$ and
  $W^{\pm}W^{\mp}$ production, the latter due to electron charge mis-measurement.
  ``Other'' background includes \ttbar{}, single top, \zjets{}, $Zh$ and $Wh$ production.
  The errors shown include statistical and systematic uncertainties.
}
\begin{center}
\scalebox{0.98}{
\begin{tabular}{l*{6}{l@{\,$\pm$\,}l}}
\toprule
& \multicolumn{2}{c}{\SReeone} & \multicolumn{2}{c}{\SReetwo}
& \multicolumn{2}{c}{\SRmmone} & \multicolumn{2}{c}{\SRmmtwo}
& \multicolumn{2}{c}{\SRemone} & \multicolumn{2}{c}{\SRemtwo} \\
\midrule
Observed events & \multicolumn{2}{l}{2} & \multicolumn{2}{l}{1}
                & \multicolumn{2}{l}{6} & \multicolumn{2}{l}{4}
                & \multicolumn{2}{l}{8} & \multicolumn{2}{l}{4} \\
SM expectation  &    6.0 &    1.2  &  2.8 &    0.8  &   3.8 &    0.9  &  2.6 &    1.1  &  7.0 &    1.3  &  1.9  &   0.7 \\
\midrule
Non-prompt      &    3.4 &    1.0  &  1.6 &    0.5  &  0.00 &   0.20  &  0.3 &    0.4  &  3.0 &    0.9  & 0.48  &  0.28 \\
$WZ$, $ZZ$      &    2.2 &    0.6  &  0.7 &    0.4  &   3.4 &    0.8  &  1.8 &    0.9  &  3.3 &    0.8  &  1.1  &   0.5 \\
$WW$            &   0.33 &   0.31  & 0.22 &   0.23  &  0.24 &   0.29  &  0.4 &    0.5  &  0.4 &    0.4  & 0.23  &  0.26 \\
Other           &   0.13 &   0.13  & 0.31 &   0.31  &  0.14 &   0.14  & 0.06 &   0.06  & 0.19 &   0.17  & 0.09  &  0.08 \\
\bottomrule
\end{tabular}}
\end{center}
\end{table*}

Depending on the final state, additional kinematic variables are used to further reduce the
background.
Requiring the pseudorapidity difference between the two leptons $\detall<1.5$
decreases the $WZ$ and $ZZ$ background.
Requirements on \metrel, defined as
\begin{equation}
\metrel = \left\{ \begin{array}{ll} \met & \mathrm{if}\ \Delta\phi>\pi/2, \\
                                   \met \sin\left(\Delta\phi\right) & \mathrm{if}\ \Delta\phi<\pi/2, \\
                 \end{array} \right.
\end{equation}
where $\Delta\phi$ is the azimuthal angle difference between $\vec{p}_\mathrm{T}^\mathrm{miss}$
and the nearest lepton or jet, reduce the $Z$ + jets and non-prompt lepton background in the $ee$ channel.
The \metrel{} is defined so as to reduce the impact on \met{} of any
potential mismeasurement, either from jets or from leptons.
The scalar sum \meff{} of the transverse momenta of the leptons, jets and the missing transverse momentum
is used to suppress the diboson background.
Requiring $\mtmax>110\GeV$, where \mtmax\ is the larger of the two
\mtW\ values computed with one of the leptons and the missing
transverse momentum, suppresses background events with one
leptonically decaying $W$ boson, whose transverse mass distribution
has an endpoint at $m_W$.

To test the non-prompt lepton and charge mismeasurement backgrounds,
validation regions are defined by applying only the number of jets \njet{} and lepton
\pt{} requirements from Table~\ref{tab:SRll} and requiring
$\mlj>90\GeV$ or $\mljj>120\GeV$.

%%%%%%%%%%%%%%%%%%%%%%%%%%%%%%%%%%%%%%%%%%%%%%%

\subsection{Background estimation}
\label{sec:ss-2l-bkg}

The irreducible background in the same-sign dilepton channel is
dominated by $WZ$ and $ZZ$ diboson production, in which both vector
bosons decay leptonically and one or two leptons do not satisfy the selection requirements, mostly the kinematic ones. These
contributions are estimated from the simulation.

Background contributions due to non-prompt leptons are estimated with
the matrix method described in Ref.~\cite{ATLAS:2L8TeV}.
It takes advantage of the difference between the efficiencies for prompt and non-prompt leptons, defined as the fractions of prompt and non-prompt preselected leptons respectively, that pass the signal-lepton requirements.
The number of events containing non-prompt leptons is obtained from these efficiencies and the observed number of events using four categories of selection with preselected or signal leptons.
The efficiencies for prompt and non-prompt leptons are derived, as a
function of \pt{} and $\eta$, for each process leading to either prompt or non-prompt
leptons using the generator-level information from simulated events. They are
then corrected for potential differences between simulation and data with correction factors measured in
control regions, as described in Ref.~\cite{ATLAS:2L8TeV}. The
contributions from each process leading to either prompt or non-prompt leptons are then used to compute a
weighted-average efficiency, where the weight for each process is
determined as its relative contribution to the number of preselected leptons
in the region of interest.

\begin{sloppypar}
Same-sign background events where the lepton charge is mismeasured are
usually due to a hard bremsstrahlung photon with subsequent asymmetric
pair production. The charge mismeasurement probability, which is
negligible for muons, is measured in data as a function of electron
\pt{} and $|\eta|$ using $Z\to e^+e^-$ events where
the two electrons are reconstructed with the same charge. The
probability, which is below $1\%$ for most of the \pt{} and $\eta$
values, is then applied to the simulated opposite-sign $ee$ and $e\mu$
pairs to estimate this background~\cite{ATLAS:ss2lstrong8TeV}.
Although any process with the $e^{\pm}e^{\mp}$ or $e^{\pm}\mu^{\mp}$
final state can mimic the same-sign signature with charge
mismeasurement, most of this background contribution is due to the
production of \zjets{} events, amounting to less than $10\%$ of the
background yield in each of the \llss{} signal regions.
\end{sloppypar}

Estimates of non-prompt lepton and charge mismeasurement background
are tested in the validation regions; the number of observed events
agrees with the expected background in all validation regions.
Figure~\ref{fig:fakeVRll} shows the distribution of \meff{} in the
validation region of the same-sign $e\mu$ channel.

The number of observed and expected events in each signal region
is reported in Table~\ref{tab:ss2l_results_blinded_bkgOnlyFit}.
Figure~\ref{fig:SRll} shows the data distributions of \meff{}, \mtmax{}, \mlj{}, and \mljj{} compared to the SM expectations in the same-sign dilepton signal regions.
No significant excess is observed over the SM background expectations in any channel.

%%%%%%%%%%%%%%%%%%%%%%%%%%%%%%%%%%%%%%%%%%%%%%%

\section{Systematic uncertainties}
\label{sec:systematics}

\begin{table*}[ht!]
\caption{\label{tab:systematics}
	Summary of the statistical and main systematic
    uncertainties on the background estimates, expressed in per cent
    of the total background yields in each signal region. Uncertainties that are not
    considered for a particular channel are indicated by a ``--''.
	The individual uncertainties can be correlated, and do
    not necessarily add in quadrature to the total background
    uncertainty.
}
\begin{center}
\scalebox{0.93}{
\begin{tabular}{lcccccc}
\toprule
& \SRlbbone & \SRlbbtwo & \SRlggone & \SRlggtwo & \SRllone & \SRlltwo \\
\midrule
Number of background events     & $6.0\pm1.3$ & $2.8\pm0.8$
                                & $1.6\pm0.4$ & $3.3\pm0.8$
                                & $16.8\pm2.8$ & $7.3\pm1.5$ \\
\midrule
Statistical                     &   9 &   7 &  22 &  23 &   7 &   7  \\
Modelling \ttbar                 &  23 &  25 &  -- &  -- & -- &  --   \\
Modelling single top             &   5 &  11 &  -- &   --&  -- &  --  \\
Modelling $Wh$, $Zh$, $\tth$     &  -- &  -- &   3 &   1 &  -- &  --  \\
Modelling $WZ$                   &  -- &  -- &  -- &  -- &  11 &  22  \\
Electron reconstruction         &   3 &   3 &   1 &   1 & $<1$& $<1$ \\
Muon reconstruction             &   1 &   1 & $<1$& $<1$&   1 & $<1$ \\
Photon reconstruction           &  -- &  -- &   4 &   5 &  -- &  --  \\
Jet energy scale and resolution &   6 &  14 &   1 &   3 &   2 &  11  \\
$b$-jet identification          &   6 &   4 &  -- &  -- &  -- &  --  \\
\mbb\ shape                     &   8 &  12 &  -- &  -- &  -- &  --  \\
Background \mgg\ model          &  -- &  -- &   5 &   7 &  -- &  --  \\
Non-prompt estimate             &  -- &  -- &  -- &  -- &  10 &  11  \\
Charge mismeasurement estimate &  -- &  -- &  -- &  -- &   2 &   3  \\
Other sources                   &   4 &   5 & $<1$&   2 &   2 &   2  \\
\bottomrule
\end{tabular}}
\end{center}
\end{table*}

Table~\ref{tab:systematics} summarises the dominant systematic
uncertainties on the total expected background yields in the six
signal regions.

\begin{sloppypar}
For the one lepton and two $b$-jets channel, theoretical uncertainties on the \ttbar\
and single-top background estimates are the most important.
They are evaluated by comparing different generators (\powheg, \mcatnlo~\cite{Frixione:2002ik,Frixione:2005vw} and \acermc)
and parton shower algorithms (\pythia\ and \herwig~\cite{Marchesini:1991ch,herwig}),
varying the QCD factorisation and renormalisation scales up and down by a factor of two, and taking the envelope of the 
background variations when using different PDF sets.
Statistical uncertainties from the data in the CRs result in uncertainties on the normalisations of the \ttbar{} and \wjets{} backgrounds, while the limited number of simulated events yields uncertainty on the shape of the background \mbb\ distributions.
The largest experimental systematic uncertainties are those on the jet energy scale~\cite{Aad:2014bia} and resolution~\cite{Aad:2012ag}, derived from a combination
of test-beam data and in-situ measurements, followed by the uncertainty on the $b$-jet identification efficiency~\cite{ATLAS-CONF-2014-004}.
The uncertainty on the $W$ boson background modelling is dominated by the uncertainty on the cross section for the production of the $W$ boson in association with heavy-flavour jets, and is reported within the ``Other sources". The $W$ boson background component is small in \lbb\ SRs, and its uncertainty is constrained by the CRs with a similar composition.
\end{sloppypar}

For the one lepton and two photons channel, the background uncertainties are dominated by 
the data statistics in the \mgg\ sidebands.
The only source of systematic uncertainty on the non-Higgs background estimate is 
the choice of \mgg\ model.
The systematic uncertainties on the Higgs background estimates are dominated by
the theoretical uncertainties on the $Wh$, $Zh$, and $\tth$ production
cross sections and the photon reconstruction.  
The main theoretical uncertainties are those on the QCD scales and the parton distribution functions~\cite{CERNYellowReport3}.
The effect of scale uncertainties on the modelling of Higgs boson production is evaluated by reweighting the simulated Higgs
boson \pt\ distribution to account for doubling and halving the
scales. The experimental systematic uncertainty from photon reconstruction is
determined with the tag-and-probe method using radiative $Z$
decays~\cite{ATLAS-CONF-2012-123}.

For the same-sign dilepton channel, the two main sources of systematic
uncertainty are related to the non-prompt lepton estimate, and to
the modelling of the $WZ$ background. The uncertainty on the non-prompt
estimate
originates mainly from the limited accuracy of the efficiency
correction factors, and on the production rate of non-prompt leptons, in
particular their $\eta$ dependence. The uncertainty on the $WZ$
background modelling is determined using a same-sign,
$WZ$-enriched sample used to validate the \sherpa{} prediction. This
validation sample is selected by requiring three leptons, two of which
must have same flavour, opposite sign, $|\mll{}-m_{Z}|<10~\GeV{}$, and
then considering only the highest-\pt{} same-sign pair.
None of the other requirements from Table~\ref{tab:SRll} are applied,
except for the lepton \pt{} and \njet{} selections.

%%%%%%%%%%%%%%%%%%%%%%%%%%%%%%%%%%%%%%%%%%%%%%%

\section{Results and interpretations}
\label{sec:results}

The event yields observed in data are consistent with the
Standard Model expectations within uncertainties in all signal regions.
The results are used to set exclusion limits with the frequentist hypothesis test
based on the profile log-likelihood-ratio test statistic and approximated with asymptotic formulae~\cite{Cowan:2010js}.

\begin{table}[b!]
\begin{center}
\caption{\label{tab:limits}
From left to right, observed 95\% CL upper limits
(\limvisobs) on the visible cross sections, the observed (\limsobs) and expected (\limsexp)
95\% CL upper limits on the number of signal events with $\pm1\sigma$
excursions of the expectation, the observed confidence level of the
background\hyp only hypothesis (\clb), and the discovery $p$-value (\pzero), truncated at $0.5$.
}
\begin{tabular}{lrrrrr}
\toprule
& \limvisobs [fb] & \limsobs & \limsexp & \clb & \pzero \\
\midrule
\SRlbbone & 0.26 & 5.3 & ${6.3}^{+3.4}_{-2.0}$ & 0.28 & 0.50 \\
\SRlbbtwo & 0.27 & 5.5 & ${5.1}^{+2.6}_{-1.4}$ & 0.56 & 0.43 \\
\midrule
\SRlggone & 0.18 & 3.6 & ${4.1}^{+2.0}_{-0.7}$ & 0.25 & 0.50 \\
\SRlggtwo & 0.34 & 7.0 & ${5.9}^{+2.0}_{-1.2}$ & 0.75 & 0.19 \\
\midrule
\SRllone  & 0.51 & 10.4 & $10.9^{+3.8}_{-3.1}$ & 0.51 & 0.50 \\
\SRlltwo  & 0.51 & 10.3 &  $8.1^{+3.3}_{-1.5}$ & 0.72 & 0.32 \\
\bottomrule
\end{tabular}
\end{center}
\end{table}

\begin{figure*}[ht!]
\begin{center}
\subfigure[One lepton and two $b$-jets channel]
	{\includegraphics[width=0.49\textwidth]{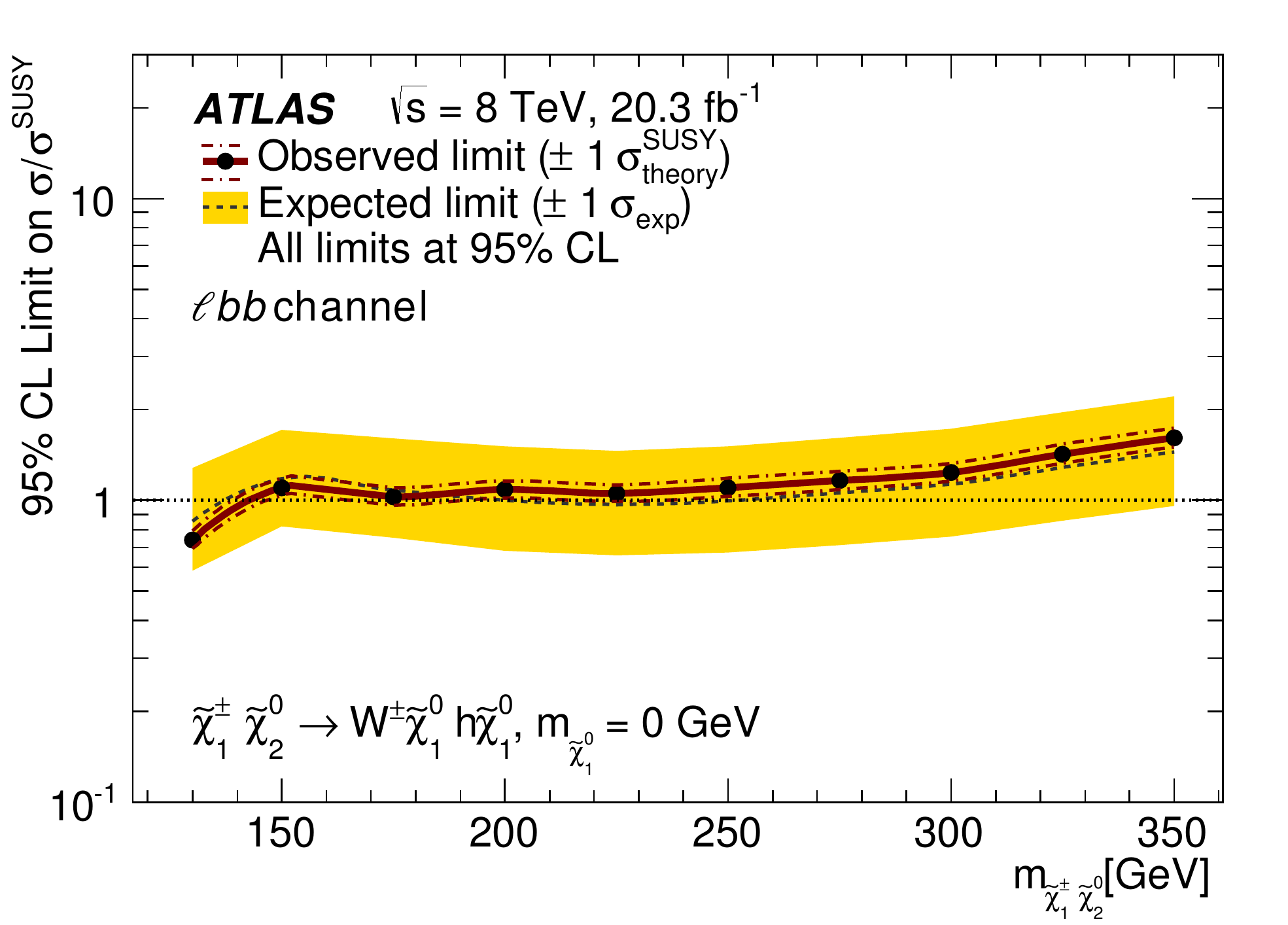}}
\subfigure[One lepton and two photons channel]
	{\includegraphics[width=0.49\textwidth]{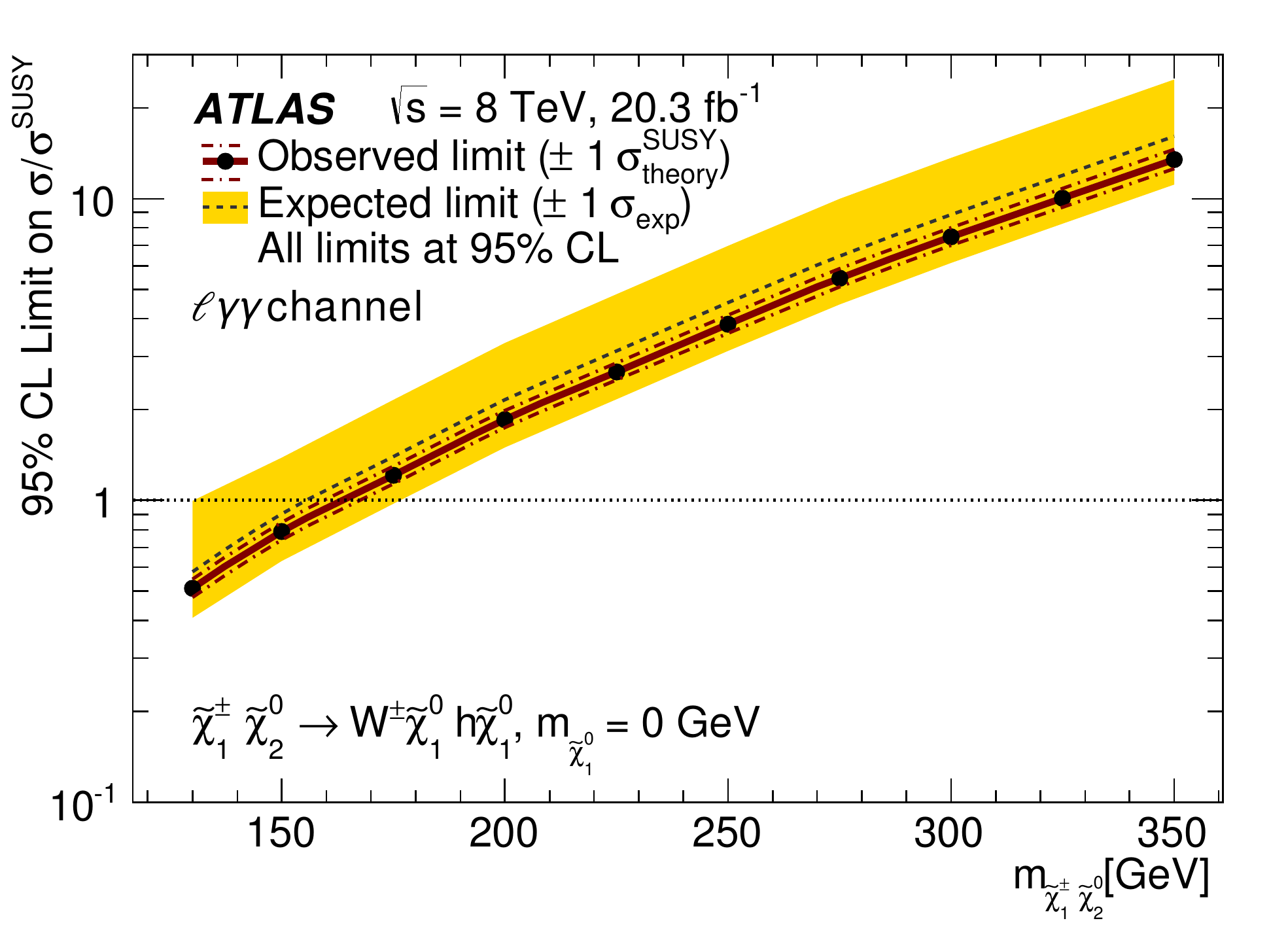}}
\subfigure[Same-sign dilepton channel]
	{\includegraphics[width=0.49\textwidth]{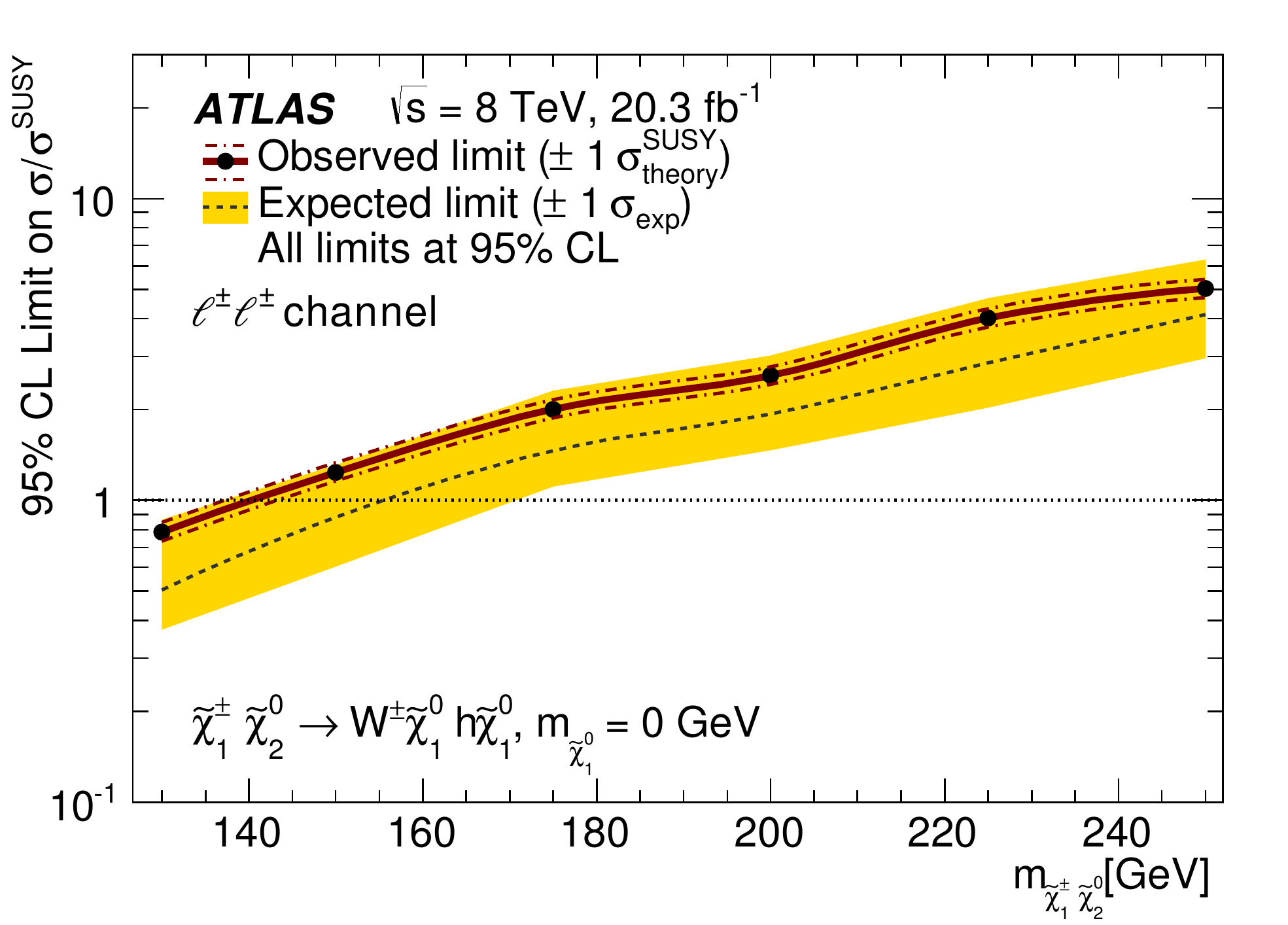}}
\subfigure[Combination]
	{\includegraphics[width=0.49\textwidth]{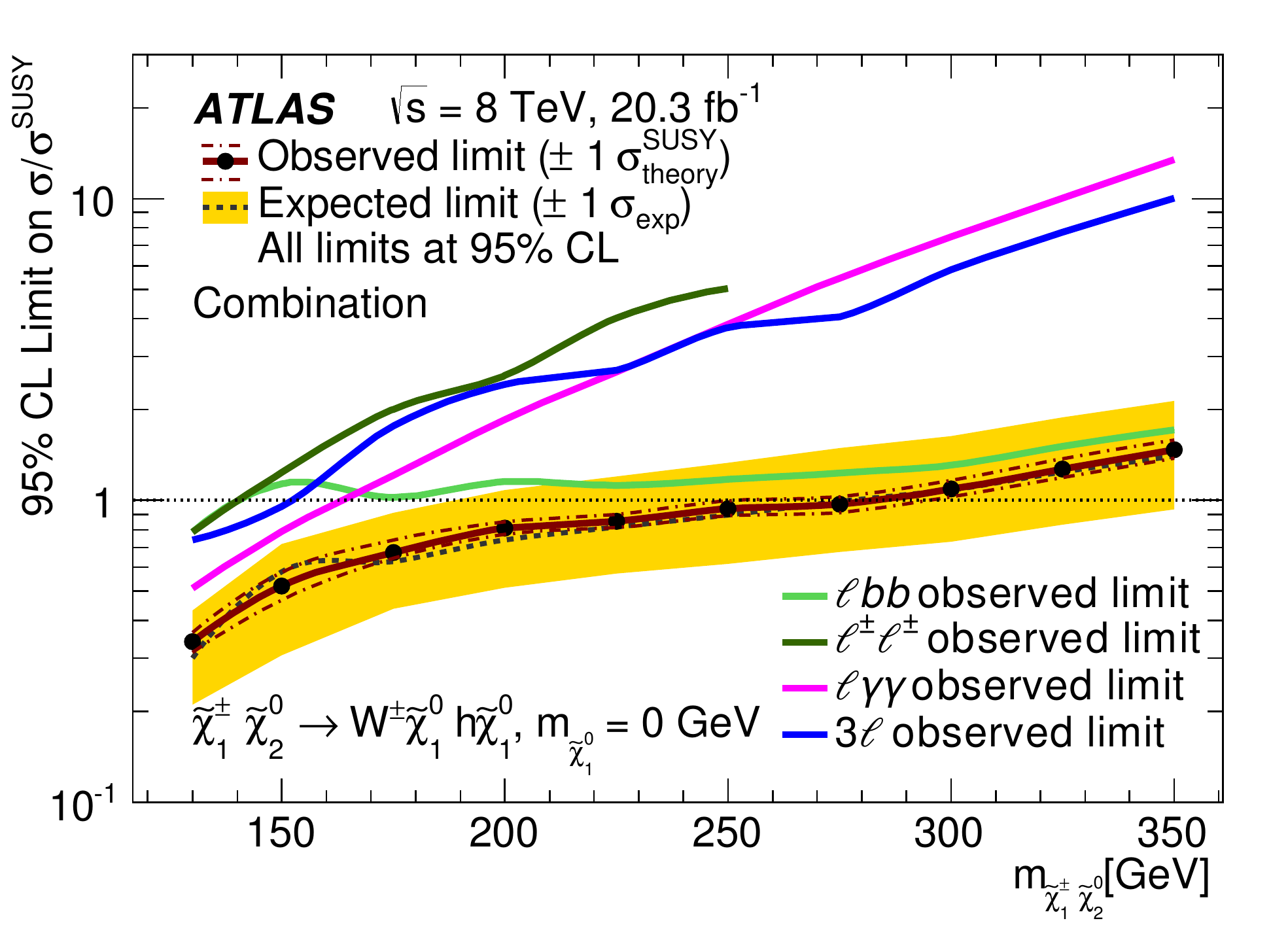}}
\caption{\label{fig:simplified1d}
	Observed (solid line) and expected (dashed line) 95\% CL upper limits
	on the cross section normalised by the simplified model prediction as a function of the common mass $m_{\chinoonepm\ninotwo}$
	for $\mNone=0$.
	The combination in (d) is obtained using the result from the ATLAS
	three-lepton search~\cite{ATLAS:3L8TeV} in addition to the three channels reported in this paper.
	The dash-dotted lines around the observed limit represent the results obtained 
	when changing the nominal signal cross section up or down by the $\pm1\sigma_{\text{theory}}^{\text{SUSY}}$ 
	theoretical uncertainty. 
	The solid band around the expected limit represents the $\pm1\sigma_{\text{exp}}$ uncertainty band where 
	all uncertainties, except those on the signal cross sections, are considered. 
}
\end{center}
\end{figure*}
\begin{figure*}[ht!]
\begin{center}
\subfigure[One lepton and two $b$-jets channel]{\includegraphics[width=0.49\textwidth]{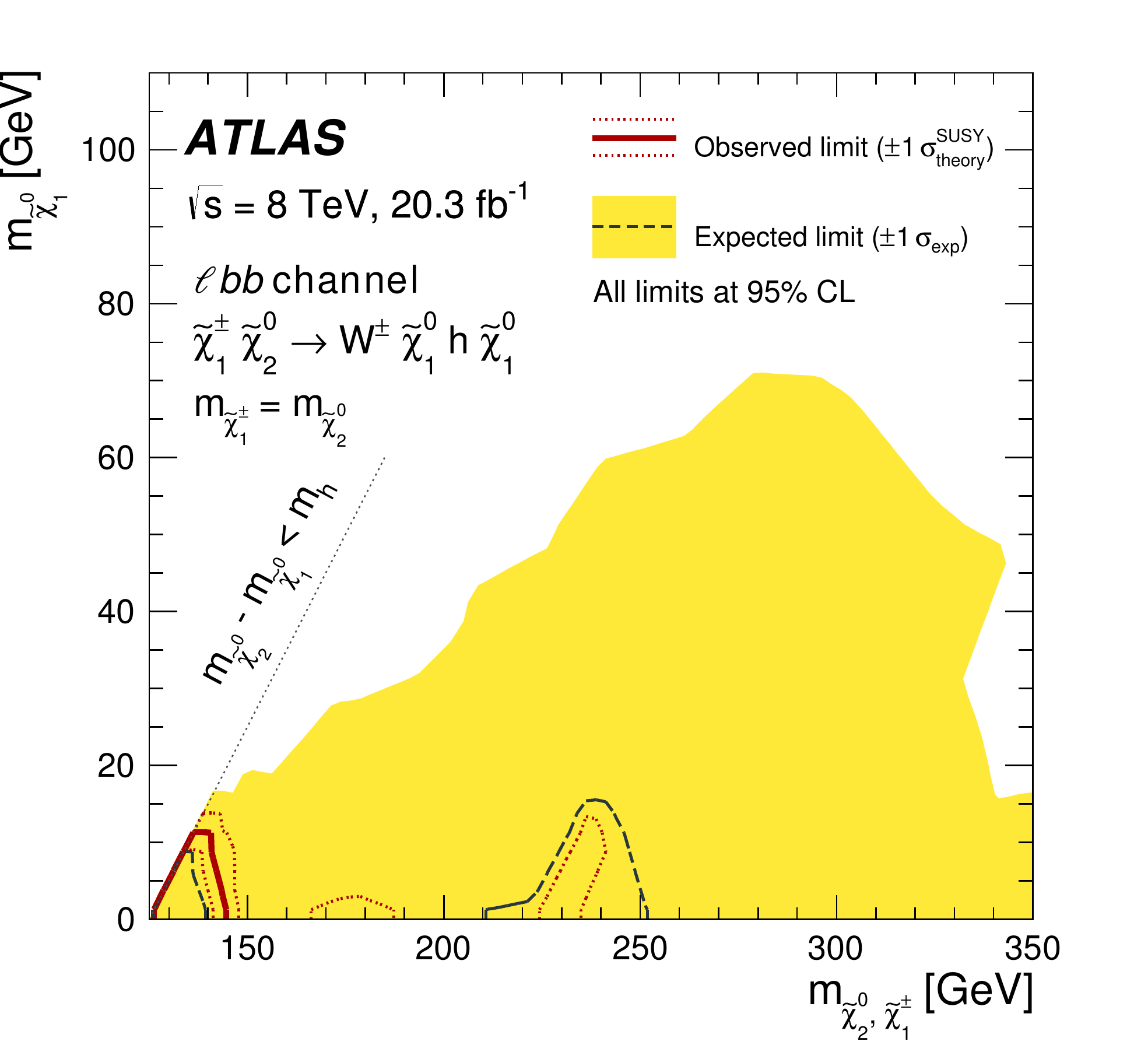}}
\subfigure[One lepton and two photons channel]{\includegraphics[width=0.49\textwidth]{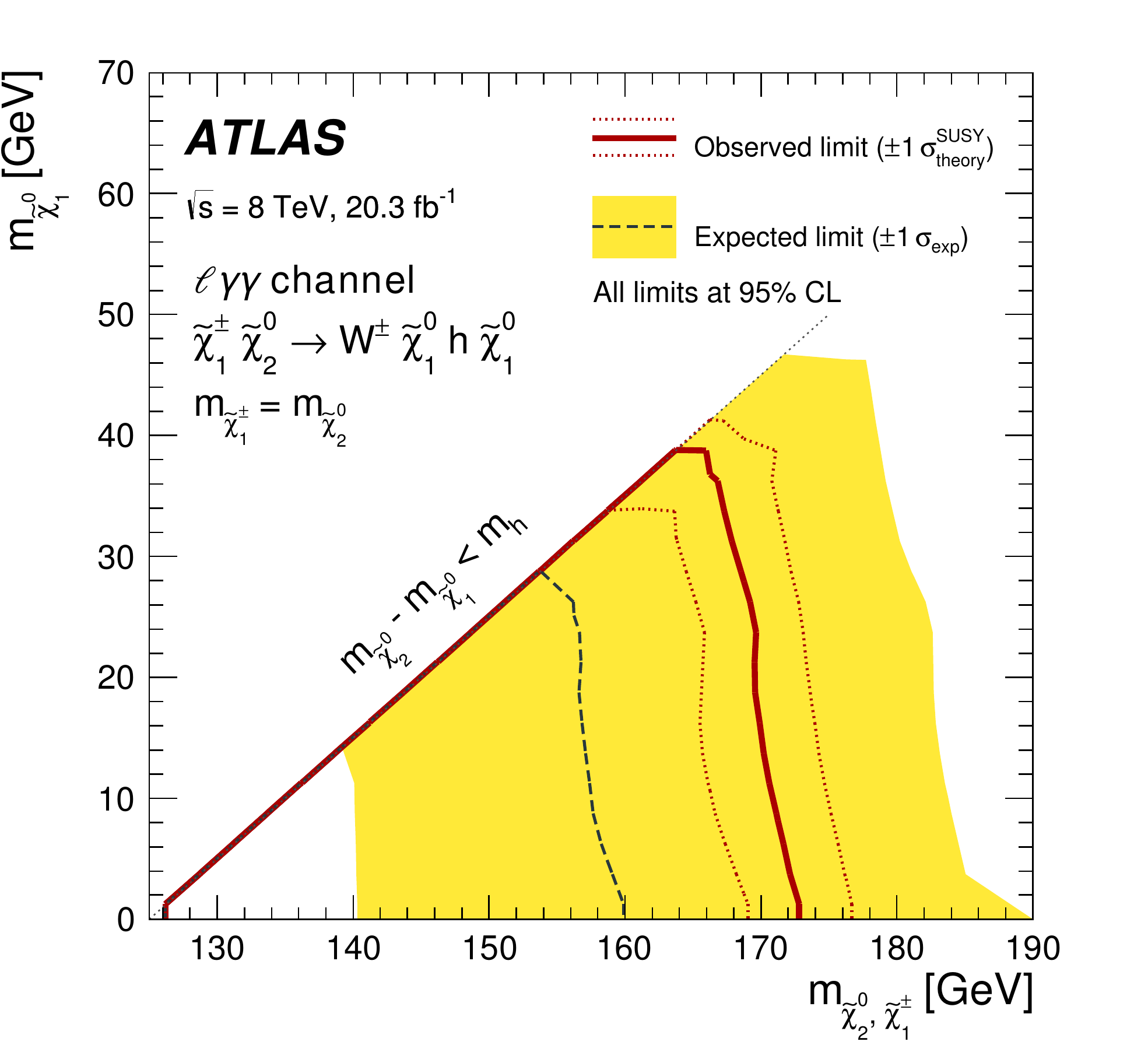}}
\subfigure[Same-sign dilepton channel]{\includegraphics[width=0.49\textwidth]{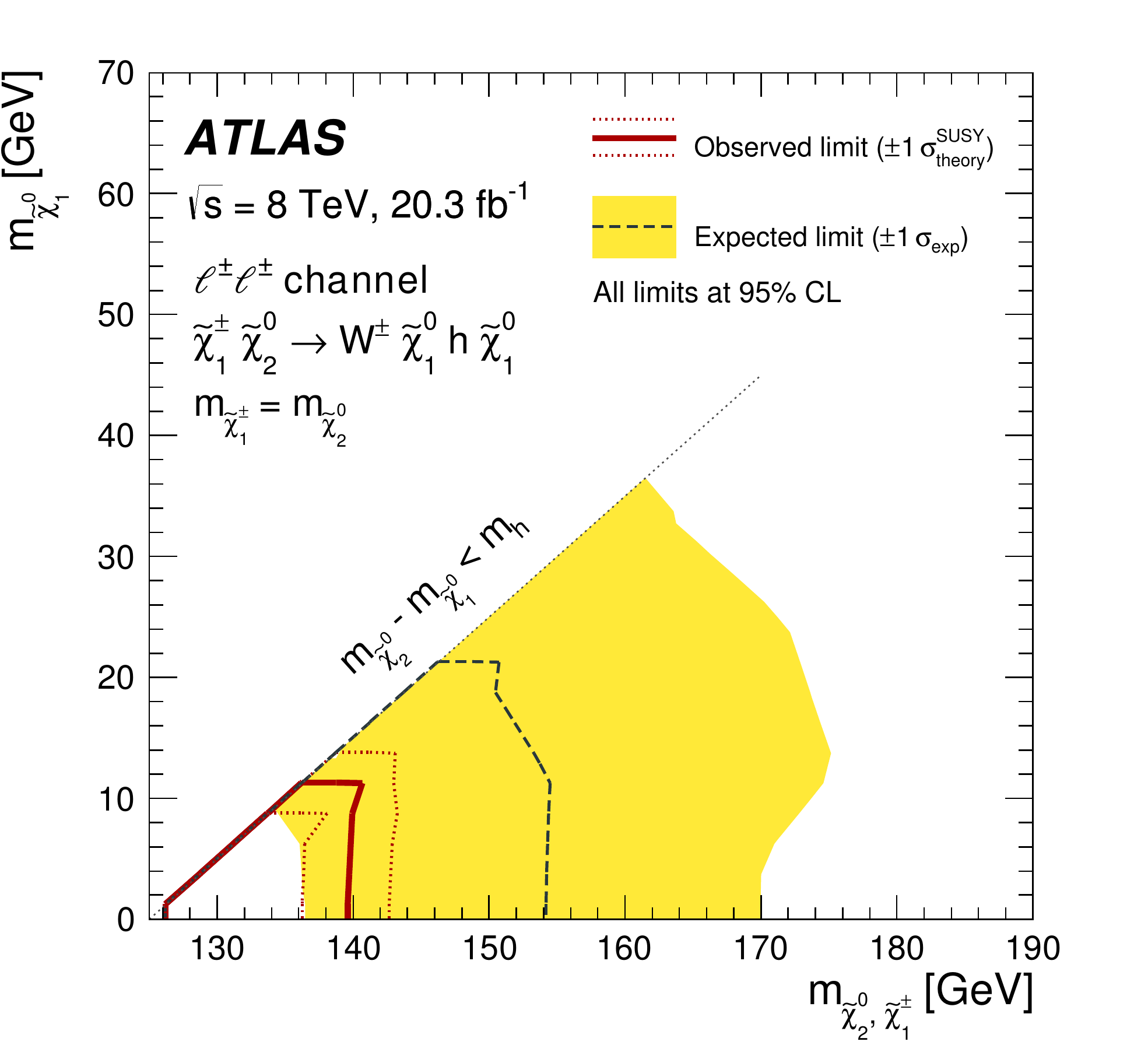}}
\subfigure[Combination]{\includegraphics[width=0.49\textwidth]{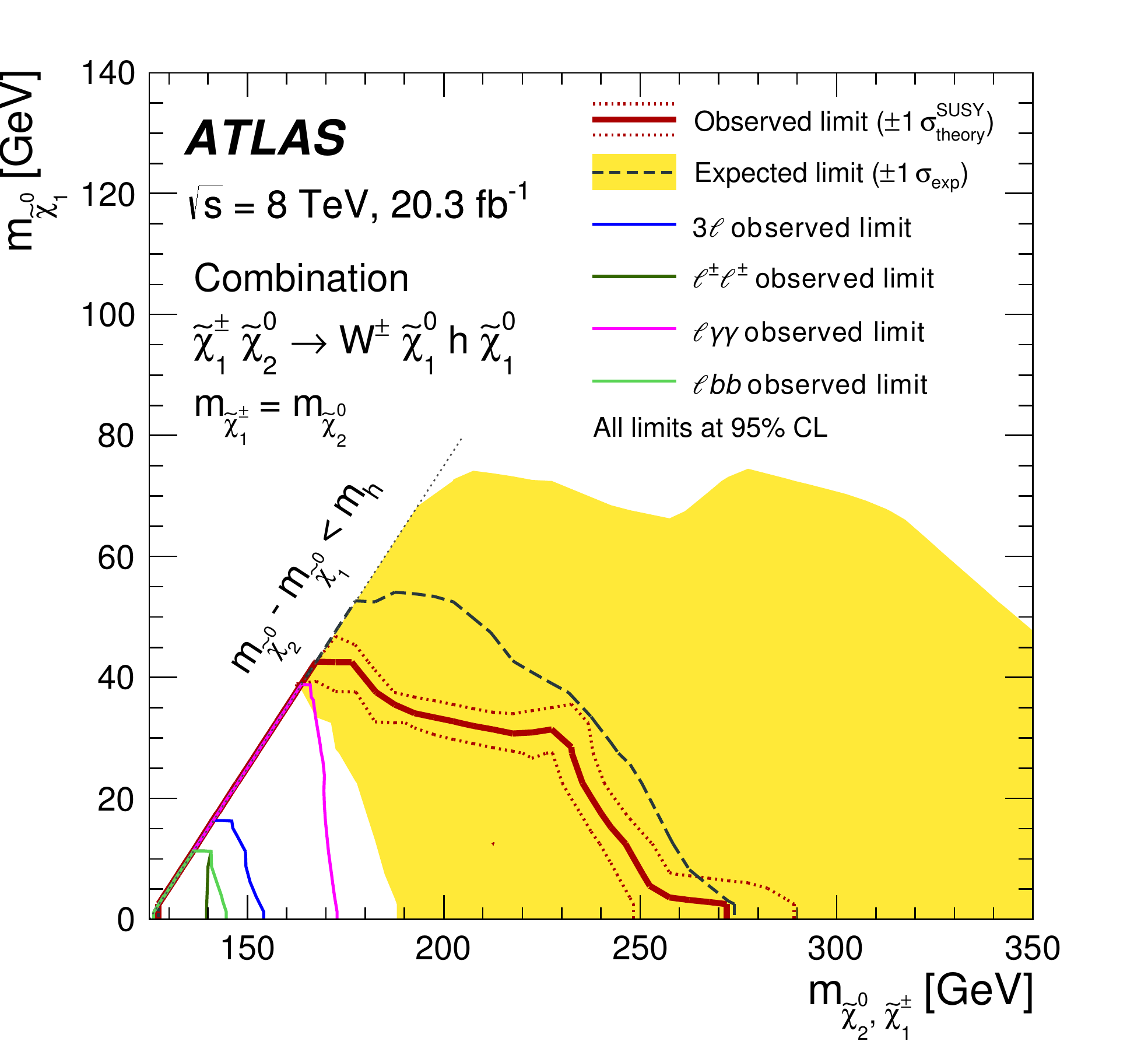}}
\caption{\label{fig:simplified} 
	Observed (solid line) and expected (dashed line) 95\% CL exclusion regions
	in the mass plane of $m_{\ninoone}$ vs.~ $m_{\ninotwo,\chinoonepm}$ in the simplified model.
	The combination in (d) is obtained using the result from the ATLAS
	three-lepton search~\cite{ATLAS:3L8TeV} in addition to the three channels reported in this paper.
	The dotted lines around the observed limit represent the results obtained 
	when changing the nominal signal cross section up or down by the $\pm 1 \sigma_{\text{theory}}^{\text{SUSY}}$ 
	theoretical uncertainty. 
	The solid band around the expected limit shows the $\pm1\sigma_{\text{exp}}$ uncertainty band where 
	all uncertainties, except those on the signal cross sections, are considered. 
}
\end{center}
\end{figure*}

Exclusion upper limits at the 95\% confidence level (CL) on the number of beyond-the-SM (BSM)
signal events, $S$, for each SR are derived using the \CLs{}
prescription~\cite{Read:2002hq}, assuming no signal yield in other signal and control regions.
Normalising the upper limits on the number of signal events
by the integrated luminosity of the data sample provides upper limits on the visible BSM cross section,
$\sigma_{\rm vis} = \sigma \times A \times \epsilon$, where $\sigma$
is the production cross section for the BSM signal, $A$ is the
acceptance defined as the fraction of events passing the geometric and
kinematic selections at particle level, and $\epsilon$ is the detector
reconstruction, identification and trigger efficiency.

Table~\ref{tab:limits} summarises, for each SR, the observed 95\% CL
upper limits (\limvisobs) on the visible cross section, the observed
(\limsobs) and expected (\limsexp) 95\% CL upper limits on the number
of signal events with $\pm1\sigma$ excursions of the expectation, the
observed confidence level (\clb) of the background\hyp only hypothesis,
and the discovery $p$-value (\pzero), truncated at $0.5$.

The results are also used to set exclusion limits
on the common mass of the \chinoonepm\ and \ninotwo\ for various
values of the \ninoone\ mass in the simplified model of
$pp\to\chinoonepm\ninotwo$ followed by $\chinoonepm\to W^\pm\ninoone$
and $\ninotwo\to h\ninoone$.
In this hypothesis test, all the CRs and SRs, including the data in the Higgs-mass windows of 
the \lbb\ and \lgg\ channels, are fitted
simultaneously, taking into account correlated experimental and
theoretical systematic uncertainties as common nuisance parameters.
The signal
contamination in the CRs is accounted for in the fit, where a single
non-negative normalisation parameter is used to describe the signal model in all
channels.

\begin{sloppypar}
Systematic uncertainties on the signal expectations stemming from
detector effects are included in the fit in the same way as for the
backgrounds.
Theoretical systematic uncertainties on the signal cross section described in Sect.~\ref{sec:mc-simulation} are not included directly in the fit.
In all resulting exclusions the dashed (black) and solid (red)
lines show the 95\% CL expected and observed limits respectively,
including all uncertainties except for the theoretical signal cross-section uncertainty. The (yellow) bands around the expected limit
show the $\pm1\sigma_{\text{exp}}$ expectations. The dotted $\pm1\sigma_{\text{theory}}^{\text{SUSY}}$ (red)
lines around the observed limit represent the results obtained when
changing the nominal signal cross section up or down by its theoretical
uncertainty, and reported limits correspond to the
$-1\sigma{}$ variation.
\end{sloppypar}

\begin{sloppypar}
Figure~\ref{fig:simplified1d} shows the
95\% CL upper limits on the signal cross section normalised by the
simplified-model prediction as a function of $m_{\ninotwo,\chinoonepm}$ for $\mNone=0$.
The sensitivity of the individual one lepton and two
$b$-jets, one lepton and two photons, and same-sign dilepton channels is illustrated in Figs.~\ref{fig:simplified1d}(a)--(c) respectively.
The corresponding limit combining all channels and the ATLAS three-lepton search is shown in
Fig.~\ref{fig:simplified1d}(d).
For $m_{\ninotwo,\chinoonepm}>250$~\GeV{} the same-sign dilepton channel is not considered.
In Fig.~\ref{fig:simplified1d}(a), the expected exclusion region
below $m_{\ninotwo,\chinoonepm}=140$~\GeV{} is largely due to
\SRlbbone, which targets models with small mass splitting between the
neutralinos, while the expected exclusion region around
$m_{\ninotwo,\chinoonepm}=240$~\GeV{} is driven by \SRlbbtwo\ designed
for larger mass splittings.
The upper limit shows slow variation with increasing $m_{\ninotwo,\chinoonepm}$ as
 the acceptance of \SRlbbtwo{} increases and compensates for the decrease of the production cross section.
 Figure~\ref{fig:simplified1d}(d) shows that in the $m_{\ninotwo,\chinoonepm}<170$~\GeV{} range all channels show similar sensitivity, while for $m_{\ninotwo,\chinoonepm}>170$~\GeV{} the one lepton and two $b$-jets channel is the dominant one.
Nevertheless, the contribution from the other channels to the combination is important to extend the excluded range significantly compared to Fig.~\ref{fig:simplified1d}(a). 
\end{sloppypar}

Figures~\ref{fig:simplified}(a)--(c) show the 95\% CL exclusion regions
in the $(m_{\ninotwo,\chinoonepm}, m_{\ninoone})$ mass plane of the simplified model obtained from the individual one lepton and two
$b$-jets, one lepton and two photons, and same-sign dilepton signal regions,
respectively.
Figure~\ref{fig:simplified}(d) shows the corresponding exclusion region obtained by combining
the three channels described in this paper with the ATLAS three-lepton
search, which by itself excludes
$m_{\ninotwo{}, \chinoonepm{}}$ up to 160~\GeV{} for
$m_{\ninoone}=0$ as seen in Fig.~\ref{fig:simplified}(d). The combination of these four independent
searches improves the sensitivity significantly, and
the 95\% CL exclusion region for $\mNone=0$ is extended to \comblimit.
The wide uncertainty bands of the expected limits in Fig.~\ref{fig:simplified} are due to the slow variation of the sensitivity with increasing $m_{\ninotwo,\chinoonepm}$ and $m_{\ninoone}$, as can also be seen in Fig.~\ref{fig:simplified1d}.
In a similar search by the CMS Collaboration~\cite{Khachatryan:2014mma}, the observed limit on $m_{\ninotwo{}, \chinoonepm{}}$ is 210 GeV for $\mNone=0$.

%%%%%%%%%%%%%%%%%%%%%%%%%%%%%%%%%%%%%%%%%%%%%%%

\section{Conclusions}	
\label{sec:conclusions}

\begin{sloppypar}
A search for the direct pair production of a chargino and a neutralino $pp\to\chinoonepm\ninotwo$
followed by $\chinopm\to \ninoone(W^{\pm}\to\ell^{\pm}\nu)$ and $\ninotwo\to \ninoone (h\to bb/\gamma\gamma/\ell^{\pm}\nu qq)$
has been performed using \lumi\ of $\sqrt{s}=8\TeV$ proton--proton collision data delivered by the Large Hadron Collider and recorded with the ATLAS detector.
Three final-state signatures are considered: one lepton and two $b$-jets, one lepton and two photons,
and two same-sign leptons, each associated with missing transverse momentum.
Observations are consistent with the Standard Model expectations.
Limits are set in a simplified model, combining these results with the three-lepton search presented in Ref.~\cite{ATLAS:3L8TeV}.
For the simplified model, common masses of \chinoonepm\ and \ninotwo\ are excluded up to \comblimit\ for a massless \ninoone.
\end{sloppypar}

%%%%%%%%%%%%%%%%%%%%%%%%%%%%%%%%%%%%%%%%%%%%%%%

\section*{Acknowledgements}

\sloppy

% Acknowledgements for papers with collision data
% Version 19-Feb-2015

% Standard acknowledgements start here
%----------------------------------------------
We thank CERN for the very successful operation of the LHC, as well as the
support staff from our institutions without whom ATLAS could not be
operated efficiently.

We acknowledge the support of ANPCyT, Argentina; YerPhI, Armenia; ARC,
Australia; BMWFW and FWF, Austria; ANAS, Azerbaijan; SSTC, Belarus; CNPq and FAPESP,
Brazil; NSERC, NRC and CFI, Canada; CERN; CONICYT, Chile; CAS, MOST and NSFC,
China; COLCIENCIAS, Colombia; MSMT CR, MPO CR and VSC CR, Czech Republic;
DNRF, DNSRC and Lundbeck Foundation, Denmark; EPLANET, ERC and NSRF, European Union;
IN2P3-CNRS, CEA-DSM/IRFU, France; GNSF, Georgia; BMBF, DFG, HGF, MPG and AvH
Foundation, Germany; GSRT and NSRF, Greece; RGC, Hong Kong SAR, China; ISF, MINERVA, GIF, I-CORE and Benoziyo Center, Israel; INFN, Italy; MEXT and JSPS, Japan; CNRST, Morocco; FOM and NWO, Netherlands; BRF and RCN, Norway; MNiSW and NCN, Poland; GRICES and FCT, Portugal; MNE/IFA, Romania; MES of Russia and ROSATOM, Russian Federation; JINR; MSTD,
Serbia; MSSR, Slovakia; ARRS and MIZ\v{S}, Slovenia; DST/NRF, South Africa;
MINECO, Spain; SRC and Wallenberg Foundation, Sweden; SER, SNSF and Cantons of
Bern and Geneva, Switzerland; NSC, Taiwan; TAEK, Turkey; STFC, the Royal
Society and Leverhulme Trust, United Kingdom; DOE and NSF, United States of
America.

The crucial computing support from all WLCG partners is acknowledged
gratefully, in particular from CERN and the ATLAS Tier-1 facilities at
TRIUMF (Canada), NDGF (Denmark, Norway, Sweden), CC-IN2P3 (France),
KIT/GridKA (Germany), INFN-CNAF (Italy), NL-T1 (Netherlands), PIC (Spain),
ASGC (Taiwan), RAL (UK) and BNL (USA) and in the Tier-2 facilities
worldwide.
%----------------------------------------------

%%%%%%%%%%%%%% References %%%%%%%%%%%%%%%%%%

\bibliographystyle{atlasBibStyleWoTitle}
\bibliography{paper}

\clearpage
\onecolumn
% ATLAS Collaboration author list
% Data extracted on 09-Feb-2015 for paper reference SUSY-2013-23
%\documentclass[11pt]{article}
%\usepackage{a4wide}\begin{document}
\begin{flushleft}
{\Large The ATLAS Collaboration}

\bigskip

G.~Aad$^{\rm 85}$,
B.~Abbott$^{\rm 113}$,
J.~Abdallah$^{\rm 152}$,
S.~Abdel~Khalek$^{\rm 117}$,
O.~Abdinov$^{\rm 11}$,
R.~Aben$^{\rm 107}$,
B.~Abi$^{\rm 114}$,
M.~Abolins$^{\rm 90}$,
O.S.~AbouZeid$^{\rm 159}$,
H.~Abramowicz$^{\rm 154}$
% ...
\bigskip
\\
% ...
$^{11}$ Institute of Physics, Azerbaijan Academy of Sciences, Baku, Azerbaijan\\
% ...
$^{85}$ CPPM, Aix-Marseille Universit{\'e} and CNRS/IN2P3, Marseille, France\\
% ...
$^{90}$ Department of Physics and Astronomy, Michigan State University, East Lansing MI, United States of America\\
% ...
$^{107}$ Nikhef National Institute for Subatomic Physics and University of Amsterdam, Amsterdam, Netherlands\\
% ...
$^{113}$ Homer L. Dodge Department of Physics and Astronomy, University of Oklahoma, Norman OK, United States of America\\
$^{114}$ Department of Physics, Oklahoma State University, Stillwater OK, United States of America\\
% ...
$^{117}$ LAL, Universit{\'e} Paris-Sud and CNRS/IN2P3, Orsay, France\\
% ...
$^{152}$ Institute of Physics, Academia Sinica, Taipei, Taiwan\\
% ...
$^{154}$ Raymond and Beverly Sackler School of Physics and Astronomy, Tel Aviv University, Tel Aviv, Israel\\
% ...
$^{159}$ Department of Physics, University of Toronto, Toronto ON, Canada\\
% ...
\end{flushleft}

%\end{document}
% Created with ./xml2latex.py

\end{document}

%%  LocalWords:  CERN Supersymmetry SUSY supersymmetric Charginos LSP
%%  LocalWords:  neutralinos eigenstates Higgs electroweak bosons CMS
%%  LocalWords:  higgsinos binos colliders charginos Tevatron LEP ZZ
%%  LocalWords:  chargino neutralino leptonically dilepton bino IP
%%  LocalWords:  centre dataset hadronically azimuthal pseudorapidity
%%  LocalWords:  microstrip LAr scintil muon toroid PDF